\documentclass[%
preprint,
amsmath,amssymb,
aps,
pra,
floatfix,
]{revtex4-1}

\usepackage{amsmath} 
\usepackage{amssymb}
\usepackage{subcaption}
\usepackage{natbib}
\usepackage{graphicx}
\usepackage{dcolumn}
\usepackage{bm}
\usepackage{tikz}
\usepackage{xcolor}
\usepackage[percent]{overpic}
\usepackage{epstopdf,epsfig}  
\usepackage{array}
\usepackage{afterpage}
\usepackage{textcomp}

\begin{document}

\title{Fluidisation of yield stress fluids under vibration}
\author{Ashish Garg$^{1,2}$}
\author{Nico Bergemann$^{2,3}$}
\author{Beccy Smith$^4$}
\author{Matthias Heil$^{2,3}$}
\author{Anne Juel$^{*1,2}$}
\affiliation{$^1$Department of Physics \& Astronomy, The University of Manchester, Oxford Road, Manchester M13 9PL, UK;}
\affiliation{$^2$Manchester Centre for Nonlinear Dynamics, The University of Manchester, Oxford Road, Manchester M13 9PL, UK;}
\affiliation{$^3$Department of Mathematics, The University of Manchester, Oxford Road, Manchester M13 9PL, UK}
\affiliation{$^4$Mondelez International, Bournville Place, Birmingham B30 2LU, UK.}

\date{\today}
\begin{abstract}
Motivated by the industrial processing of chocolate, we study experimentally the fluidisation of a sessile drop of yield-stress fluid on a pre-existing layer of the same fluid under vertical sinusoidal oscillations. We compare the behaviours of molten chocolate and Carbopol which are both shear-thinning with a similar yield stress but exhibit very different elastic properties. We find that these materials spread when the forcing acceleration exceeds a threshold which is determined by the initial deposition process. However, they exhibit very different spreading behaviours: whereas chocolate exhibits slow long-term spreading, the Carbopol drop rapidly relaxes its stress by spreading to a new equilibrium shape with an enlarged footprint. This spreading is insensitive to the history of the forcing. In addition, the Carbopol drop performs large-amplitude oscillations with the forcing frequency, both above and below the threshold. We investigate these viscoelastic oscillations and provide evidence of complex nonlinear viscoelastic behaviour in the vicinity of the spreading threshold. In fact, for forcing accelerations greater than the spreading threshold, our drop automatically adjusts its shape to remain at the yield stress. We discuss how our vibrated-drop experiment offers a new and powerful approach to probing the yield transition in elastoviscoplastic fluids.
\end{abstract}

\maketitle

\section{Introduction}

Yield-stress fluids are amongst the most common and versatile
materials in everyday life, ranging from food and healthcare products
to concrete and crude oil. Their relevance extends from the
geophysical scale, e.g., mudslides and lava flow \cite{balmforth:2014} to the
microscale where, e.g., self-assembled peptide gels are used as a
vehicle for targeted drug delivery \cite{saini}. These materials all exhibit
a transition from solid to fluid-like behaviour when subjected to
stresses exceeding the material's yield stress but they may have
widely different mesostructures, e.g. microgels such as Carbopol, suspensions, emulsions \cite{bonn2}. Industrial processing of yield-stress materials often involves their temporary fluidisation under mechanical stress in order to facilitate moulding, e.g., of chocolate or concrete, and bottling and packaging, e.g. cosmetics and food products \cite{coussot2005rheometry}. In chocolate manufacture, vibration is routinely imposed at key stages along the production line in order to shape or ensure uniform distribution of the material \cite{chevalley:1975,bergemann:2015}. 
However, vibrational parameters such as amplitude, frequency and direction of forcing are typically selected empirically. Thus, there is scope to optimise this process by gaining a fundamental understanding of the mechanics of fluidisation under vibration. In this paper, we study a configuration relevant to chocolate production lines, namely the fluidisation under vertical oscillations of a drop of yield-stress fluid initially at rest on a thin layer of the same material. 

Thus far, the study of yield-stress fluid flows has focussed mostly on uniform flows in rheometric and conduits geometries, flows in non-uniform vessels of varying cross-section, transient flows such as gravity currents and spin-coating processes and some elongational flows involving droplet formation \cite{coussot:2014}. Thus, literature on the fluidisation of yield-stress fluids under vibration is scarce. Flow enhancement and liquefaction of pastes and complex fluids in vibrated pipes have been explored \cite{deysarkar:1981,deshpande:2001} and the effect of vertical oscillations on a layer of complex fluid has generated growing interest in recent years. For example, a signature of elasticity was identified in the viscoelastic Faraday problem using a wormlike micelle solution \cite{ballesta:2005}. Convection-driven heap formation and cracking was observed in dense slurries of bronze microspheres \cite{schleier:2001} while persistent holes were found to form in vibrated layers of cornstarch or glass sphere suspensions \cite{merkt:2004}. 
In yield-stress materials, \citet{shiba:2007} observed convective motion in drops of shear-thinning polymer gels when driven by vertical oscillations above a critical acceleration, which resulted in organised patterns of periodic deformation of the drop. The onset of this convective motion is governed by a constant ratio, marginally greater than unity, of the vibrational stress due to inertia to the yield stress of the material. \citet{wolf:2015} generated regular patterns of holes in a more extended layer of Carbopol microgel subject to vertical vibration, which persisted when the forcing was discontinued, returning the layer to a solid state. These spatial features suggest that different modes of deformation of the free-surface may be accessed depending on the vibrational parameters. The mechanics of yielding under vibration, however, remain poorly explored.

In the last decade or so, intense rheological research coupled to flow
visualisation has provided new insight into yielding from a microscopic
point of view \cite{bonn2}, and shown that the increase of stresses in
the material towards the yield threshold leads to its restructure,
indicated by a plethora of intriguing phenomena such as unsteady
flows, creep, transient shear-banding and slip at solid boundaries
\cite{balmforth:2014, divoux:2011, coussot:2014}. Understanding the detailed
mechanics of yielding is key to the development of accurate constitutive
models, which can in turn be applied to predict flow in a variety of
practically-relevant configurations. Such models may be derived from
the microscopic dynamics \cite{fielding:2020} or the relationship between
stress, strain and rate of strain, formulated at a macroscopic scale to
capture the mechanics observed experimentally. Arguably the most
successful constitutive model to date is the Saramito model
\cite{saramito:2009}, which treats the material as a linear viscoelastic
solid (with a characteristic relaxation time) for stresses below yield and an elastoviscoplastic fluid with a
Herschel-Bulkley viscosity above yield. Direct experimental
evidence of the role of viscoelasticity around the yield point has
only emerged recently \cite{balmforth:2014, bonn:2017, ewoldt:2008, luu:2009, piau:2009} and
views diverge on whether a linear-to-nonlinear viscoelastic
transition occurs prior to reaching the yield threshold \cite{fernandes:2017}. 

In this paper, we investigate fluidisation of yield-stress materials under vibration by comparing the behaviours of molten chocolate and Carbopol microgel, which are both shear-thinning yield-stress fluids with the same yield stress but exhibit widely different elastic properties. We perform experiments on drops initially at equilibrium under gravity on a layer of the same fluid, by oscillating the substrate vertically. Below the onset of spreading, the chocolate drop remains at rest in the frame of reference of the oscillating substrate whereas the Carbopol drop exhibits periodic viscoelastic deformations at the frequency of the forcing. Although both chocolate and Carbopol drops fluidise for similar values of the forcing acceleration, we observe qualitatively different spreading behaviours. Whereas in the chocolate drop an initial phase of rapid spreading is followed by slow, long-term dynamics, the Carbopol drop swiftly spreads to a new equilibrium shape about which it continues to oscillate viscoelastically at the forcing frequency, featuring large-amplitude stretching and compression of the drop. This results in much reduced spread of the Carbopol drop compared to the chocolate drop. 

We also find that the threshold for the onset of spreading is dependent on the process by which the initial drop shape is obtained. However, for each subsequent increase in forcing acceleration, the vertically oscillated drop automatically adjusts its shape by spreading in order to relieve its stress so that it remains at the yield threshold. This implies that compared to the standard procedures used for oscillatory rheometry \cite{hyun:2011}  -- the most widely used methodology for the characterisation of non-Newtonian fluids -- our experimental setup provides unique access to the effects of nonlinear viscoelasticity near the yield threshold. This is difficult in oscillatory shear rheometry because for yield-stress fluids that are capable of undergoing large viscoelastic deformations, the periodic shear strain imposed on the sample can be absorbed in viscoelastic or plastic deformations. Hence the oscillatory nature of the forcing ensures that even plastic deformations are effectively reversed over each cycle of the forcing. Our measurements suggest a reduction in the storage modulus of Carbopol at the yield point with increasing forcing, which concurs with rheometric measurements \cite{fernandes:2017,varges:2019,digiuseppe:2015}. We demonstrate that our vibrated-drop setup offers a new and powerful experimental approach to probing the yield transition in elastoviscoplastic fluids.

The outline of the paper is as follows. The experimental methods used to create drops, oscillate them and image them are described in \S \ref{sec:exp}. Results are presented in \S \ref{results} where we compare the behaviours of chocolate and Carbopol drops under vibration in \S \ref{sec:dropspreading}. We show that both materials fluidise at a critical forcing acceleration in \S \ref{sec:threshold}. In \S \ref{sec:timehistory} we demonstrate that the viscoplastic spreading of Carbopol is insensitive to its forcing history and in \S \ref{sec:viscoelastic}, we investigate the viscoelastic oscillations of the Carbopol drop. We summarise our findings in \S \ref{disc} and discuss the potential for our vibrated-drop setup to act as a rheometer providing elastoviscoplastic material properties near the yield point.


\section{Experimental methods}
\label{sec:exp}

\begin{figure}[!hbt]
\includegraphics{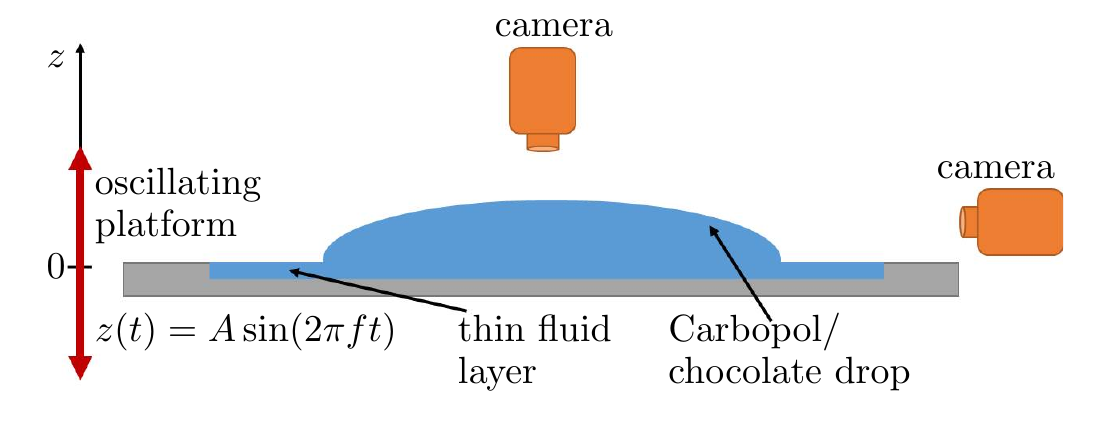}
 \caption{\small Schematic diagram of the experimental setup. The vertical displacement of the oscillating platform from its equilibrium position is given by $z(t) =A \sin(2 \pi f t)$.}
 \label{fig:platform_setup}
\end{figure}

A schematic diagram of the experimental system is shown in figure~\ref{fig:platform_setup}. A drop of yield-stress fluid is at rest on a thin layer of the same material contained within a circular trough on a rigid, horizontal substrate. We oscillate the substrate sinusoidally so that the vertical  displacement of the platform is $z(t) = A \sin(2\pi f t)$, where $A$ and $f$ are the amplitude and frequency of oscillation, respectively, and we monitor the evolution of the drop with side and top-view cameras. 

\subsection{Materials, drop deposition and substrate preparation} \label{sec:deposition of droplet}

We studied two yield-stress fluids: (i) a commercially available, tempered milk chocolate with a fat content of 30\% (in molten form), and (ii) Carbopol Ultrez-21 microgels (Lubrizol) with different concentrations.

\begin{table}[hbt]
  \begin{center}

  \begin{tabular}{lcccccc}
  \hline 
      Materials  & $\rho$~(kg $m^{-3}$)  & $\tau_y$~(Pa) &  $K$~(Pa s$^n$) & $n$ & $G$~Pa& \\
     \hline 
     Chocolate    &1,247&  $36.1 \pm 0.6$ & ~~$8.7 \pm 0.02$~ & $1 $ & $\sim 44,000$ \\
       Carbopol (2.2 g/L) &1,020  & $35 \pm 5$ & ~~$8.2 \pm 2.0$~ & $0.49 \pm 0.02$ & 350\\
       Carbopol (3 g/L)   & 1,030& $60 \pm 6$ & ~~$12 \pm 3$~ & $0.50 \pm 0.02$  & 450\\
       Carbopol (6 g/L)  &1,050 & $120 \pm 10$ & ~~$17 \pm 4$~ & $0.51 \pm 0.02$ & 700 \\
\hline 
  \end{tabular}
  \caption{\small Density and rheological properties of the commercially available, tempered milk chocolate with 30\% fat content, and Carbopol microgels of different concentrations. The yield stress $\tau_y$, the power index $n$ and the constant $K$ were obtained by fitting a one-dimensional Herschel-Bulkley model of the form $\tau = \tau_y + K|\dot{\epsilon}|^n$ ($\tau \ge \tau_y$) \cite{herschel:1926} to flow curves measured experimentally. In the case of chocolate, the coefficients were obtained from a two-parameter fit to the Bingham model ($n=1$) \cite{bergemann:2018b}. $G$ is the elastic shear modulus of the material obtained  with constant-shear rheometry for $\tau  < \tau_y$ in chocolate by \cite{bergemann:2018b}. For Carbopol we list the values measured by \cite{digiuseppe:2015, garg:2020} with oscillatory shear rheometry in the limit of an approximately elastic response ($\tau \ll \tau_y$). \label{Table1}}
 \end{center}
\end{table}

The chocolate was tempered using a Model Prima tempering machine (FBM, Italy) which delivered chocolate at its outlet at a temperature of $(28.2\pm0.3)^\circ$C. The temper level was determined prior to deposition of each drop onto the substrate, using an EXOTHERM 7400 temper meter (Systech Analytics SA, 170 Switzerland). 
The tempered chocolate has been extensively characterised by \citet{bergemann:2018b}, who performed rheological measurements using a Brookfield R/S-Plus (SST) rheometer with a four-bladed vane spindle rotating in a large-diameter cylindrical glass beaker. In that study, shear was applied cyclically by successively increasing and decreasing the angular velocity or the torque with constant rate. Values of angular velocity and torque were converted into shear rate $\dot{\epsilon}$ and stress $\tau$, respectively. The dynamic yield stress $\tau_y$ was determined as the value of decreasing stress at which the fluid came to rest. For $\tau \ge \tau_y$, the resulting flow curve of stress as a function of strain rate was found to be best fitted by a two-parameter Bingham model (see table \ref{Table1} for parameter values). For $\tau<\tau_y$, the material deformed elastically, and the measured elastic modulus was found to be moderate ($G \sim 700$~Pa) prior to initial fluidisation, but increased to $G \sim 44,000$~Pa in subsequent cycles, indicating the mesoscopic reorganisation of the material following fluidisation. 

 We prepared three Carbopol solutions with different concentrations by adding Carbopol Ultrez-21 powder in concentrations of 2.2~g/L, 3~g/L and 6~g/L to a 0.021 Molar NaOH solution ($pH=12.3$) at the laboratory temperature of 20~$^o$C. The powder was initially dissolved by vigorous stirring with an electric whisk for a few minutes, followed by intermittent manual stirring over a period of $2-3$ hours in order to achieve homogeneous solutions. The resulting Carbopol solutions had $pH \simeq 8.0 \pm 0.5$.  They were each placed in an airtight container and rested for at least 3 days before they were used in the experiments. 
 Prior to experiments, we performed rheometric tests analogous to those reported by \citet{bergemann:2018b} for tempered chocolate, which we present in Appendix \ref{AppA}. They were carried out repeatedly over a period of several months and did not reveal measurable aging of the material, consistent with previous studies \cite{varges:2019,digiuseppe:2015}. The flow curve was accurately captured by a Herschel-Bulkley constitutive relation 
 and the values of the fitting parameters are listed in Table \ref{Table1}. 
 
The tempered chocolate and the 2.2~g/L Carbopol solution have a similar yield stress of $\tau_y\simeq 35$~Pa. Table \ref{Table1} also gives the values of the elastic shear modulus below yield. For Carbopol, we list values from \cite{digiuseppe:2015,garg:2020} obtained with oscillatory shear rheometry in the limit of an approximately elastic response for different Carbopol grades, which range from $G\sim 350$~Pa to 700~Pa. This means that the elastic shear modulus of chocolate is approximately two orders of magnitude larger than that of the 2.2~g/L Carbopol solution. 

\begin{figure}[!t]
	\includegraphics[width=\textwidth]{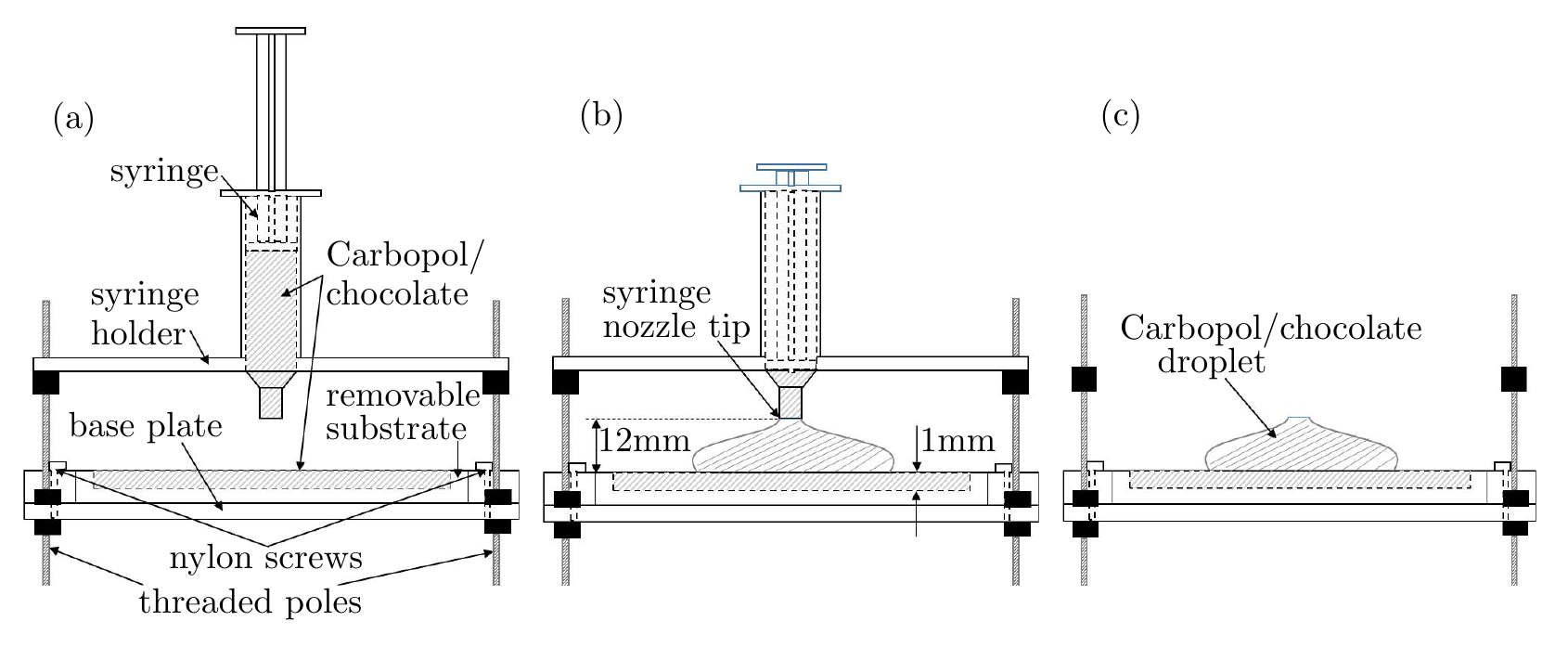}
	\caption{\small Schematic diagram of the substrate assembly and illustration of the procedure used to deposit the drop.}
	\label{fig:processofdeposition}
\end{figure}

A schematic diagram of the experimental apparatus used to prepare the drop is shown in figure \ref{fig:processofdeposition}(a).
A square Perspex substrate plate of width 105 mm was secured to a Perspex base plate with three finger-tight nylon screws. It featured a circular trough with a diameter of $60.5$~mm and a depth of $1$~mm where a uniform film of yield-stress material of depth $0.9\pm0.1$~mm was deposited prior to experimentation by initially overfilling the trough and then slowly scraping off the excess fluid with a square-edged ruler at an angle of around $30^\circ$, taking care to minimise the washboard instability at the free surface \cite{hewitt:2012}. 
The drop of yield-stress fluid was deposited onto this thin layer by extrusion from a standard 10 mL plastic syringe with a nozzle enlarged to an inner diameter of $8\pm0.05$~mm. To ensure reproducible deposition, we placed the syringe inside a vertical, tightly-fitting, removable holder which was rigidly mounted on three vertical, threaded poles above the substrate plate. We ensured the axisymmetry of the deposited drop by levelling the entire assembly
to less than $0.5^\circ$ from the horizontal using a digital inclinometer. 

The syringe was partially filled with $4.5\pm 0.3$~mL of yield-stress fluid, taking care to avoid trapping any air bubbles, and placed in the holder (figure \ref{fig:processofdeposition}(a)) so that the end of the nozzle was 12~mm above the substrate. The plunger of the syringe was then
slowly pushed down manually for approximately 3 seconds to empty the syringe. The material spread onto the substrate yielding a drop at equilibrium under gravity following deposition (figure \ref{fig:processofdeposition}(b)); see \S \ref{slump}. Finally, the deposition assembly holding the syringe was removed from the apparatus, which was then ready for the vibration experiments (figure \ref{fig:processofdeposition}(c)).  

\subsection{Oscillatory forcing}
\label{sec:oscillatory}

\begin{figure}[!hbt]
	\includegraphics[width=\textwidth]{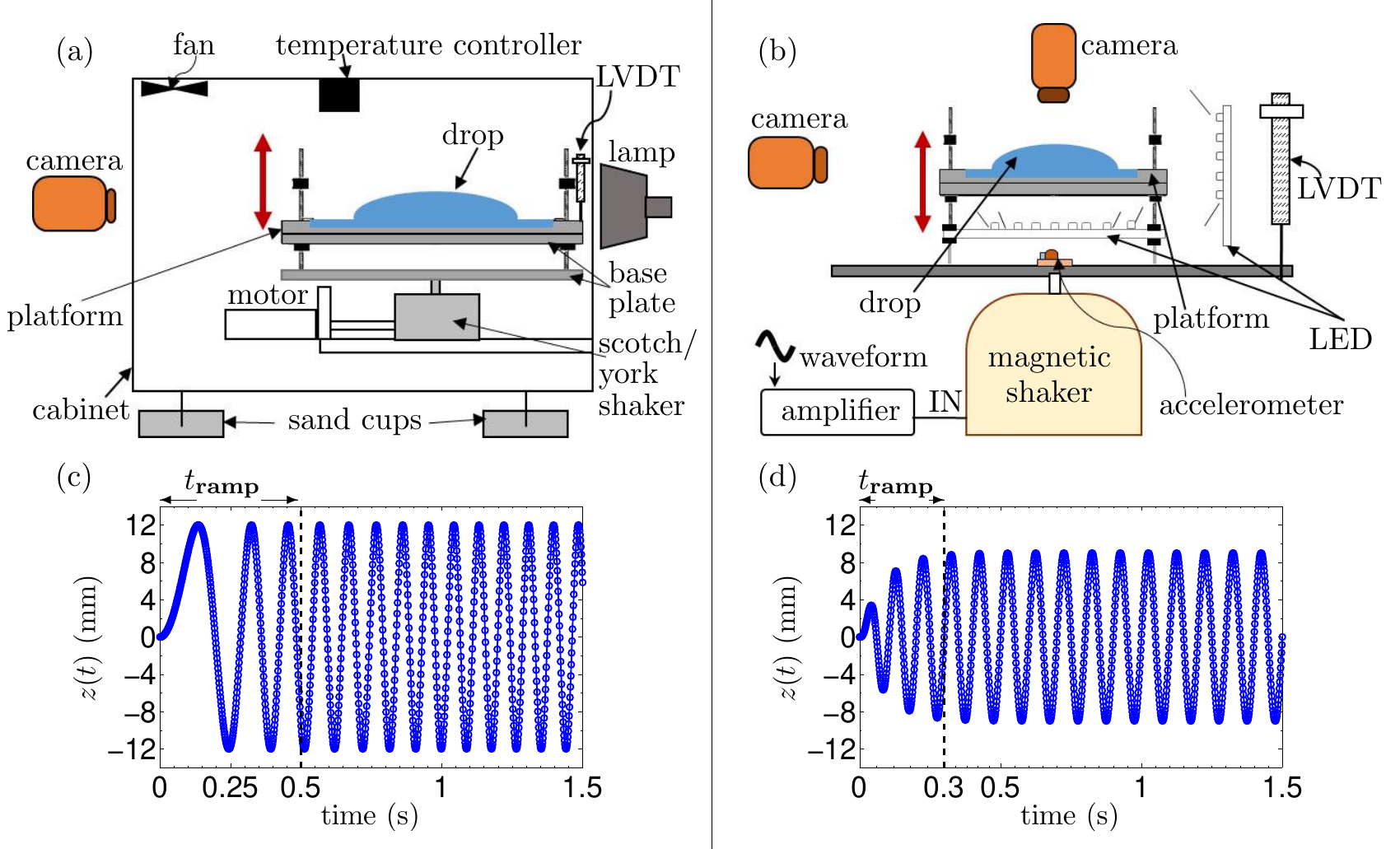}
\caption{\small (a, b) Schematic diagrams of the two experimental rigs used to impose the vertical, oscillatory motion of the substrate. (a) Rig~1: brushless DC motor attached to a Scotch yoke mechanism. (b) Rig~2: Electromagnetic shaker. (c, d) Time-series of the vertical position of the platform started from rest showing the different ramp-up procedures available in the two systems, which are applied over a time-interval $t_{\rm ramp}$ for (c) Rig~1, (d) Rig~2.}
\label{fig:two_set-ups}
\end{figure}

We used two experimental rigs to vibrate our sessile drop, as shown schematically in figure \ref{fig:two_set-ups}. In Rig~1 the rotation of a DC brushless servo motor (McLennan M644CI500L) is converted to sinusoidal translation motion via the Scotch yoke mechanism (figure \ref{fig:two_set-ups}(a)), while in Rig~2 oscillations are applied with an electromagnetic shaker (LDS V455) (figure \ref{fig:two_set-ups}(b)). Both pieces of apparatus were bolted onto heavy steel tables with adjustable feet positioned onto sand contained in four steel cylinders, in order to minimise the transfer of vibration through the floor. Rig~1 was limited to moderate frequencies and accelerations ($f \le 10$~Hz, $A(2 \pi f)^2 \le 5g$, where $g$ is the acceleration of gravity), which were sufficient to perform the experiments on tempered chocolate. In contrast, frequencies and accelerations of up to $f=50$~Hz and $A(2\pi f)^2 =12.5g$ were applied in experiments with Carbopol using Rig~2. 
\vspace{0.5em}

\subsubsection{Rig~1: brushless DC motor with Scotch yoke mechanism}

Rig~1(figure \ref{fig:two_set-ups}(a)) was originally designed for the tempered chocolate experiments. Thus, the apparatus was enclosed in a temperature-controlled cabinet which was held at $29 \pm 0.5^o$C during experiments with tempered chocolate and at the laboratory temperature of $20.2 \pm 0.5^o$C during experiments with Carbopol.

The substrate assembly described in \S \ref{sec:deposition of droplet} was adjustably mounted on three threaded poles 20~mm above a square aluminium plate (thickness 8~mm and width 250~mm) which was coupled to the vertical shaft of a Scotch yoke mechanism. The Scotch yoke was driven by a computer-controlled brushless DC motor via a 5:1 planetary gearbox (HPE70-S-5:1).
For a constant rotation rate $2 \pi f$ of the motor shaft, the Scotch yoke produced a sinusoidal displacement of the platform in the vertical direction. The frequency $f$ could be incremented during an experimental run by varying the rotation rate of the motor but the amplitude was set manually in the range $6\;\mathrm{mm} \le A \le 12.5\;\mathrm{mm}$ (to within of $\pm 0.1$~mm) prior to switching the motor on. The oscillations of the platform were calibrated to the input signal by measuring the displacement of the platform with a Linear Voltage Differential Transformer (LVDT, Solartron, Mach 1) mounted onto the platform, see figure \ref{fig:two_set-ups}(a). Accelerometer data (PCB Piezotronics) recorded at the centre of the substrate plate indicated negligible harmonic content 
over the entire range of frequency investigated ($f \leq 10$~Hz). A camera (Pulnix TM-6740CL) was mounted on the laboratory wall to capture side view images of the drop which was backlit by a halogen lamp, see \S \ref{sec:visualisation}. 

\subsubsection{Rig~2: electromagnetic shaker}

To reach the larger platform accelerations required for the Carbopol experiments, we used a permanent magnet shaker (LDS, V455) as shown in figure \ref{fig:two_set-ups}(b). The substrate assembly described in \S \ref{sec:deposition of droplet} was adjustably mounted on three threaded poles 20~mm above a circular aluminium plate (thickness 8~mm and diameter 250~mm) which was coupled to the shaker. 
The drop was lit uniformly by two LED panels, a horizontal panel below the substrate assembly and a vertical panel mounted on the laboratory wall. It was imaged using cameras in side-view (Pulnix TM-6740CL) and top-view (DALSA HM1400). 

The shaker was driven by a sinusoidal waveform generated in LabVIEW which was converted to an analog output with a data acquisition board (NI PCI-6221) and amplified to the power required by the shaker with a linear amplifier (LDS, PA 1000L). As in Rig~1, the oscillations of the platform were calibrated to the input signal  with an LVDT (Solartron, Mach 1) mounted onto the platform, see figure \ref{fig:two_set-ups}(b). The maximum displacement amplitude imposed by the shaker in our experiments was $A=9.5$~mm and the maximum frequency was $f=50$~Hz. 

The waveform of the platform oscillation was sampled with an accelerometer (PCB Piezotronics) mounted at the centre of the platform. Fast Fourier Transform (FFT) of the accelerometer signal gave a percentage power of the first harmonic relative to the fundamental mode of less than 0.2~\% for all amplitudes if $f\geq 30$~Hz, increasing to approximately 1.5~\% for $A=2$~mm and 3.2~\% for $A=8$~mm at $f=10$~Hz. Thus for $f \le 10$~Hz, we only used this rig to perform experiments at small amplitudes $0.15\;\mathrm{mm} \le A \le 5.5\;\mathrm{mm}$ which were not accessible in Rig~1 and for which the harmonic power ratio was less than 2\%. 


\subsubsection{Start-up and shut-down of the forcing} \label{start-up effects}

In each rig, start-up and shut-down protocols were required to allow the system to reach its set oscillation parameters from rest or when the peak acceleration was increased or decreased from a previous setting. Typical time series of the vertical displacement of the platform starting from rest are shown in figures \ref{fig:two_set-ups}(c, d) for the brushless DC motor (Rig~1) and the shaker (Rig~2), respectively. The shut-down procedure is not shown but was the same as the start-up procedure in reverse. In each graph, the duration of the start-up phase is indicated by $t_{\rm ramp}$. In Rig~1, the amplitude is constant and the rotation rate increases with an imposed acceleration of either $20 \pi$~rad~s$^{-2}$ (chocolate experiments) or $40 \pi$~rad~s$^{-2}$ (Carbopol experiments) towards its set frequency $f$ so that the ramping time is either $t_{\text{ramp}}= 0.1f$~s or $t_{\text{ramp}}= 0.05f$~s. In figure \ref{fig:two_set-ups}(c), the imposed acceleration is $40 \pi$~rad~s$^{-2}$ and $f=10$~Hz so that $t_{\text{ramp}}= 0.5$~s. 
In Rig~2, a fixed frequency was imposed by the input signal but it was necessary to ramp up the platform oscillation amplitude in order to avoid damaging the shaker. We followed the manufacturer's specifications and linearly increased the voltage to the amplifier until the value associated with the chosen parameter values ($f$, $A$) was reached, which led to ramp-up times $t_\text{ramp}\le0.8$~s. 
In figure \ref{fig:two_set-ups}(d), the oscillating platform reaches $A=9$~mm in $t_\text{ramp}=0.3$~s.


\subsection{Image analysis} 
\label{sec:visualisation}

We used the top-view camera to check the axisymmetry of the drops with a resolution of 13.4 pixels/mm. The mean ratio of minimum to maximum diameter of the drops presented in this paper was $D_\mathrm{min}/D_\mathrm{max} = 0.986 \pm 0.004$. In some experiments, we inclined the camera by $30^\circ$ to capture the free-surface deformation of the drop at rates between 20 and 40 frames per second (fps); see figure \ref{fig:Carbopolthreshold} in \S \ref{sec:threshold}. 
The side-view camera was levelled and aligned with the substrate in its reference  position, which was the maximum and mean height in Rigs~1 and 2, respectively.  Vertical cross-sections of the drop were imaged with a resolution of 11.2~pixels/mm. The maximum difference in viewing angle when the platform was at its maximum and minimum heights was less than $1^\circ$, which led to differences in the measured height of a sessile drop of $<1$~\%. However, we measured the height of the drop relative to the height of the substrate in the same vertical plane of view to mitigate the effects of perspective arising due the variation in the viewing angle. The side-view camera recorded images at between 100 and 140 fps and we synchronised it to the drive to record images stroboscopically at a specific phase of the platform oscillation cycle. 

\begin{figure}
	\includegraphics{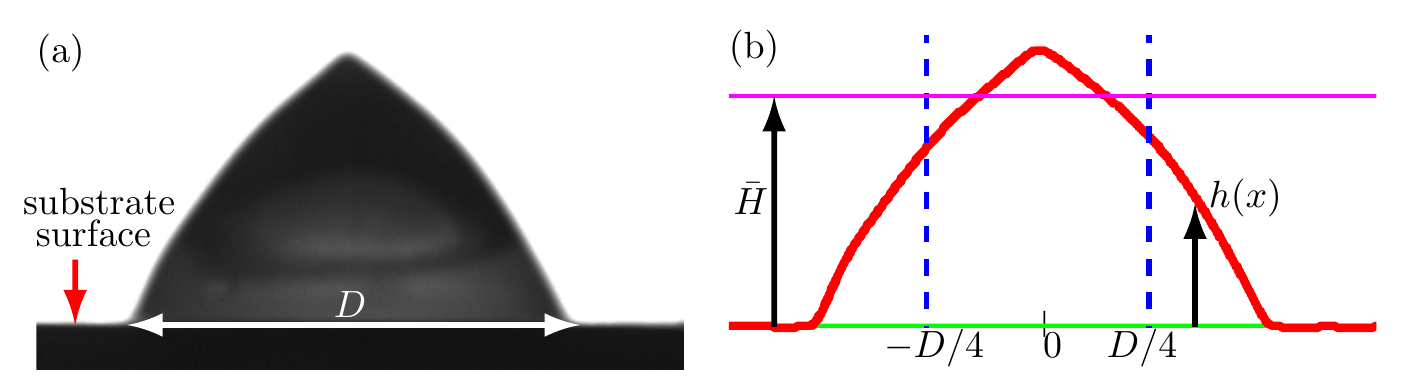}
\caption{\small Processing of side-view images: (a) raw image; (b) contour extracted using MATLAB. $h(x)$ is the local height of the drop, $D$ its diameter and the mean height $\bar{H}$ is the integral of the drop height for $-D/4\le x \le D/4$.}
\label{fig:low_higha}
\end{figure}

A typical side-view image of a Carbopol drop is shown in figure \ref{fig:low_higha}(a). A MATLAB routine based on the Canny algorithm was used to determine the contour of the drop and the position of the horizontal top surface of the platform, see figure \ref{fig:low_higha}(b). We measured the diameter $D$ of the drop $5$ pixels above the top surface of the platform and defined the mean drop height $\bar{H}$ as the local drop height $h(x)$ integrated over the central half of the drop,
\begin{equation} \label{hmea}
\bar{H} = \frac{2}{D}\int_{\frac{-D}{4}}^{\frac{D}{4}} h(x) dx.    \nonumber
\end{equation}
This metric was chosen in order to reduce the influence on the drop height of the central material thread resulting from the deposition process, and allow comparisons between chocolate and Carbopol, see figure \ref{fig:sessile}.
By analysing series of successive images, we obtained time series of the vertical substrate displacement $z(t)$, the drop height $\bar{H}(t)$ and the drop diameter $D(t)$. The error from image analysis was on the order of a pixel, and thus in the range from $1.0\%$ to $2.5\%$ for the drop height and from $0.5\%$ to $1.5\%$ for the substrate displacement and drop diameter. 

\subsection{Initial drop shape}
\label{slump}

\begin{figure}[!hbt]
	\includegraphics{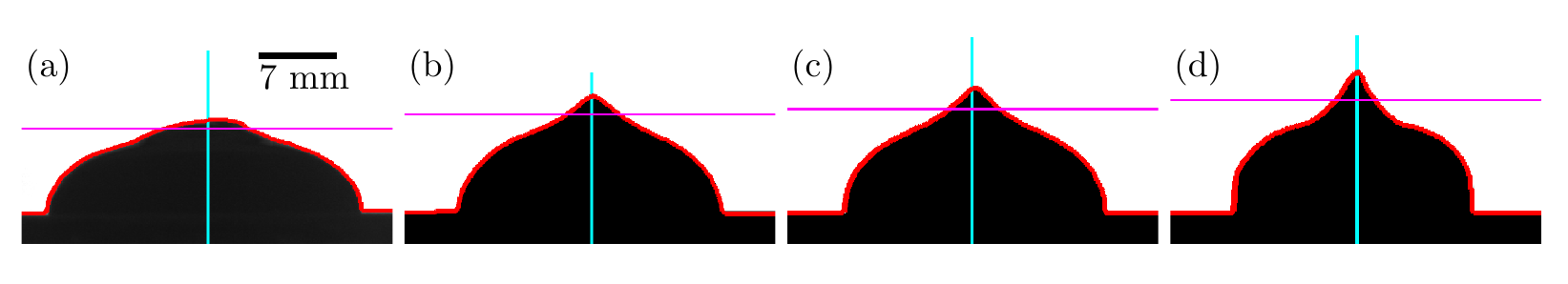}
	\caption{\small Initial shape of the droplet (a) chocolate, (b) Carbopol 2.2~g/L, (c) Carbopol 3.0~g/L, (d) Carbopol 6.0~g/L. The processed images show the drop interface and the platform edge (red line). The pink line is the mean height of the drop averaged over the central half of the drop. The mean height and diameter of each drop are: (a) $H_0=8.1 \pm 0.2$~mm, $D_0=30 \pm 0.6$~mm; (b) $H_0=9.8 \pm 0.3$~mm, $D_0=23.8 \pm 0.6$~mm; (c) $H_0=10.3 \pm 0.3$~mm, $D_0=22.4 \pm 0.6$~mm; (d) $H_0=11.1 \pm 0.3$~mm, $D_0=19.6 \pm 0.6$~mm. \label{fig:sessile}}
\end{figure}

Figure \ref{fig:sessile} shows the drop shapes associated with each material listed in table \ref{Table1}. We denote the mean height and diameter of the drops measured prior to imposing the oscillatory forcing by $\bar{H}_0$ and $D_0$, respectively. The chocolate drop (a) exhibits a more pronounced slump than the 2.2g/L Carbopol drop (b), which has a similar yield stress but an approximately 20\% smaller density (see table \ref{Table1}). The reduced spread of Carbopol drops at higher concentrations is due primarily to the associated increase in yield stress. The shape of the chocolate drop profile differs significantly from the Carbopol drops (b--d) in that the thread of material that is pinched off upon removal of the syringe collapses and merges into the drop, whereas for Carbopol it remains erect.

\section{Results}
\label{results}

\subsection{Drop spreading under vibration}
\label{sec:dropspreading}

\begin{figure}[!t]
	\includegraphics{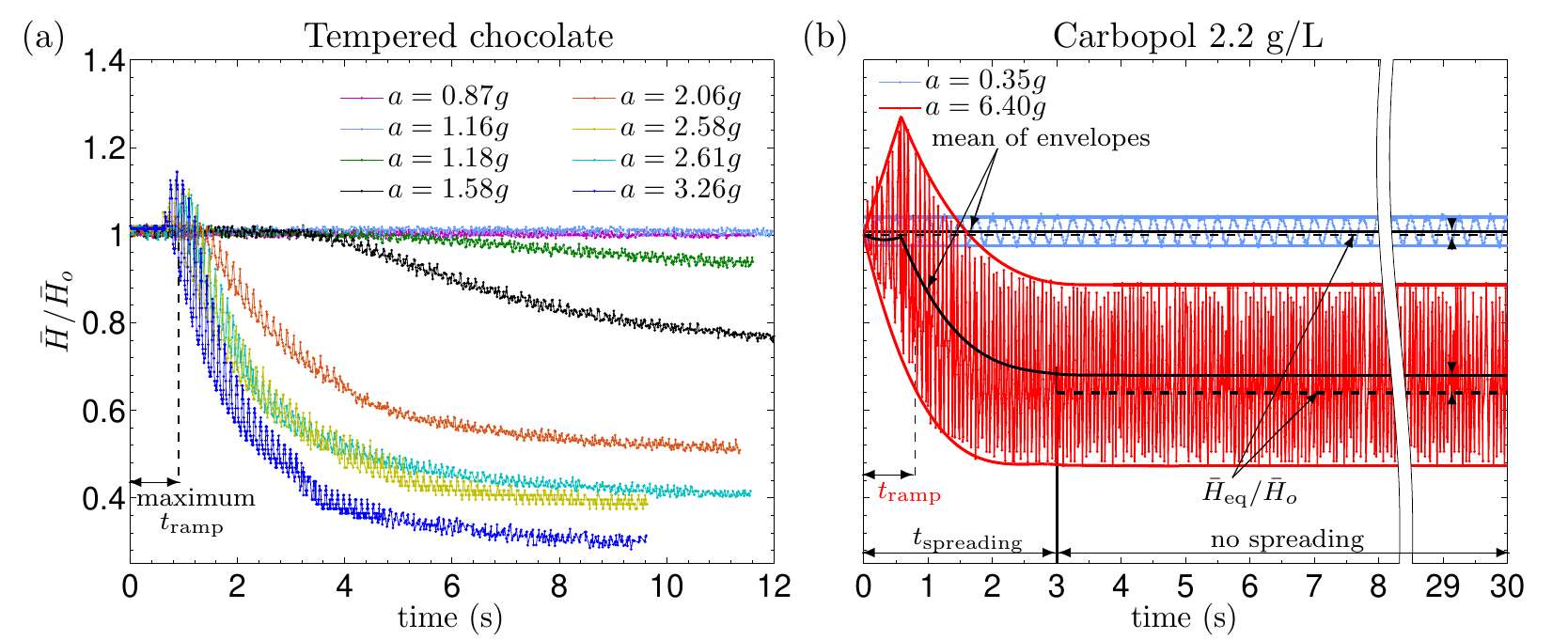}
	\caption{\small Evolution of the drop height of (a) chocolate, and (b) Carbopol (2.2~g/L) for different values of acceleration, where the forcing was imposed at $t=0$. Data for chocolate is shown for $a=0.87g$ ($A=6$~mm and $f=6$~Hz), $a=1.16g$ ($A=8$~mm and $f=6$~Hz), $a=1.18g$ ($A=6$~mm and $f=7$~Hz),
	$a=1.58g$ ($A=8$~mm and $f=7$~Hz), $a=2.06g$ ($A=8$~mm and $f=8$~Hz),
	$a=2.58g$ ($A=8$~mm and $f=9$~Hz), $a=2.61g$ ($A=10$~mm and $f=8$~Hz) and $a=3.26g$ ($A=10$~mm and $f=9$~Hz). Data for Carbopol is for $a=0.35g$ (blue: $A=5.5$~mm and $f=4$~Hz) and $a=6.40g$ (red: $A=5.5$~mm and $f=17$~Hz). In (b), the maximum extension and compression are traced with solid lines of the same colour (envelopes). The mean of the envelopes is shown with a solid black line. The equilibrium heights $\bar{H}_\mathrm{eq}/\bar{H}_0$ measured after interruption of the forcing in both experiments are denoted by dashed black horizontal lines.}
	\label{fig:dynamics_chocolate_carbopol_above}
\end{figure} 

 Figure \ref{fig:dynamics_chocolate_carbopol_above} shows the time evolution of the drop height for (a) chocolate and (b) Carbopol (2.2~g/L) from the instant substrate oscillations were imposed ($t=0$) for different values of the forcing acceleration. For low forcing acceleration ($a \le 1.16g$), the chocolate drop remains essentially undeformed, with $\bar{H}/\bar{H}_0$ varying by less than $\pm 0.5\%$. This is in contrast with the Carbopol drop which deforms periodically with the substrate oscillations, with amplitudes on the order of 10\% of the initial drop height for $a=0.35g$. Once the forcing was interrupted, the Carbopol drop returned to its initial height (dashed horizontal line) to within the experimental resolution, thus indicating the absence of permanent deformation (spreading). The lack of spreading of the drops under weak forcing suggests that in this regime both chocolate and Carbopol behave as viscoelastic solids. The absence of measurable oscillations in the chocolate drop is consistent with its larger elastic modulus; see table \ref{Table1}.

\begin{figure}[!t]
 	\includegraphics{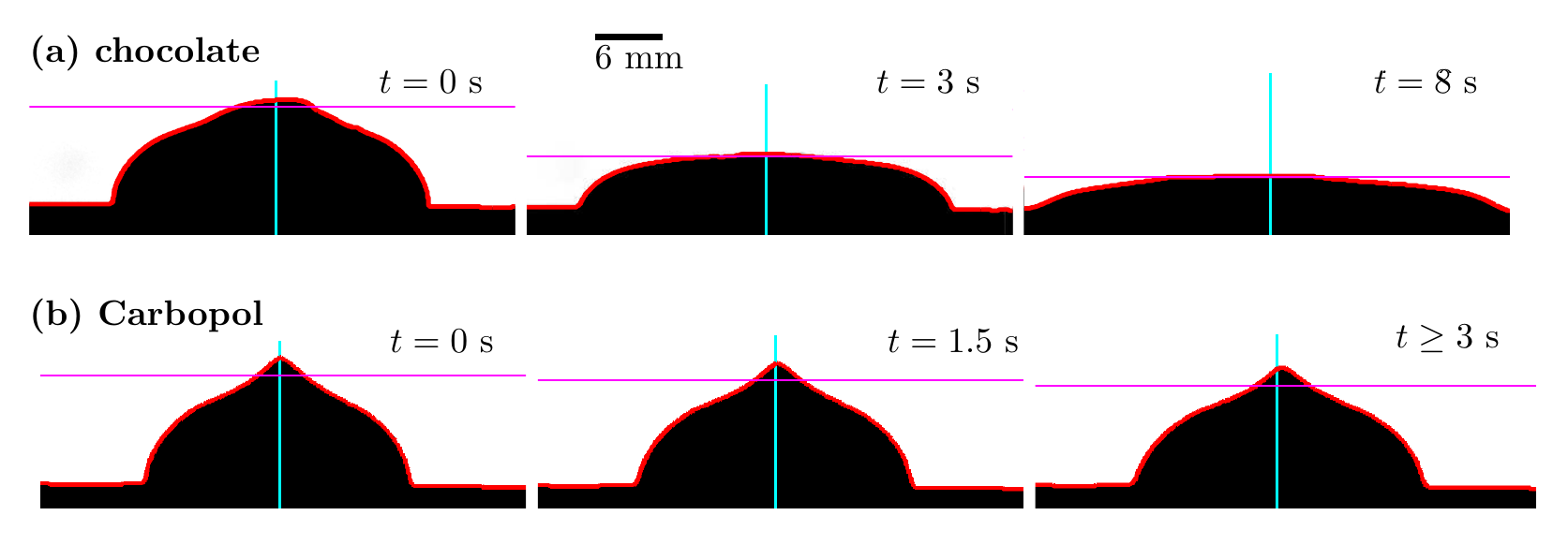}
 	\caption{\small Snapshots of chocolate (a) and Carbopol (2.2~g/L) (b) drops spreading under vertical oscillations, taken with the substrate is in its mean position. The processed images are similar to those shown in figure \ref{fig:sessile}. (a) Chocolate:
 		$a=3.26g$, $A = 10$~mm, and $f = 9$~Hz. (b) Carbopol
 		2.2~g/L: $a = 3.25g$, $A = 5.6$~mm, $f = 12$~Hz.}
 	\label{fig:images_chocolate_carbopol}
 \end{figure}

  In chocolate, the decrease of $\bar{H}/\bar{H}_0$ with time for $a \ge 1.18g$ indicates the spreading of the chocolate drop. For $a=1.18g$ this reduction becomes apparent only for $t>6$~s which is much larger than the start-up time $t_\mathrm{ramp} = 0.7$~s. Conversely for the largest acceleration ($a=3.26g$), spreading occurs much sooner (from $t\simeq1$~s, which is on the order of the start-up time, $t_\mathrm{ramp} = 0.9~s$). For all values of $a$, the spreading is sufficiently slow that the drop does not reach a constant height by the end of the experimental recording of up to $12$~s. However, for $a\ge 2.06g$ this slow long-term spreading is preceded by a rapid decrease in drop height during which the drop oscillates with the frequency of the drive. The amplitude of these viscoelastic oscillations diminishes as the drop spreads and falls below experimental resolution between $t=4$ and $5$~s. Their transient nature is likely associated with the mesoscopic reorganisation of the chocolate following fluidisation. This was already observed by \citet{bergemann:2018b} who found a considerable increase (by a factor of 6) of the pre-yield elastic modulus of the material following the first cycle of rheometric yield measurements. In the vibrated drop experiment a larger number of the faster substrate oscillation cycles during which the drop is momentarily fluidised may be required to achieve a similar effect. This would be consistent with the observed progressive reduction in oscillation amplitude. For the largest value of $a=3.26g$, an apparent initial increase in mean height is visible because the extension of the drop is larger than its compression during each cycle of the oscillation. Stroboscopic side-view images of this experiment are shown in figure \ref{fig:images_chocolate_carbopol}(a) to highlight the spreading of the fluidised drop. The drop height is reduced by $50\%$ at $t=3$~s, and more than $70\%$ at $t=8$~s. Moreover, the shape of the initial drop rapidly evolves to a smooth profile of uniformly positive curvature.
 
A typical example of the spreading of a Carbopol drop is shown in figure \ref{fig:dynamics_chocolate_carbopol_above}(b) for $a=6.40g$. The drop develops large-amplitude shape oscillations during the ramp-up of the forcing, before rapidly spreading, while continuing to oscillate, to reach a time-periodic state of constant mean value (indicated by the solid black line) for $t \gtrapprox 3$~s. This time-periodic state is similar to that observed at low forcing ($a=0.35g$), albeit with significantly larger oscillation amplitude. It indicates that the stress in the drop has decreased below $\tau_y$ so that the drop behaves as a viscoelastic solid. Upon stopping the forcing, the drop relaxes to a new equilibrium height indicated by the horizontal dashed line. Figure \ref{fig:dynamics_chocolate_carbopol_above}(b) shows that, as for the chocolate drop, the mean height of the drop is larger than its equilibrium height because the amplitude of the extensional deformation is larger than its compressional deformation. This feature is also visible in the ramp-up oscillations. We performed experiments for 13 values of acceleration up to $a=8g$, corresponding to frequencies in the range $8 \; \mathrm{Hz}\le f \le 23$~Hz, and found that although the extent of the spreading increased with increasing forcing acceleration, the duration of the spreading remained approximately constant with $t_{\rm spreading}=3.2 \pm 0.8$~s.

Figure \ref{fig:images_chocolate_carbopol} shows that spreading is considerably reduced in Carbopol compared with chocolate. For $a = 3.25g$, the final height reduction of the Carbopol drop (after interruption of the forcing) is $12\%$, while in chocolate it exceeds $70\%$ by $t=8$~s. A further increase of the substrate acceleration to $a=6.40g$ (not shown) results only in a final height reduction of the Carbopol drop of $36\%$. Moreover, the Carbopol drop retains its characteristic pointed shape which suggests only partial fluidisation in contrast with chocolate.


\begin{figure}[!ht]
	\includegraphics[]{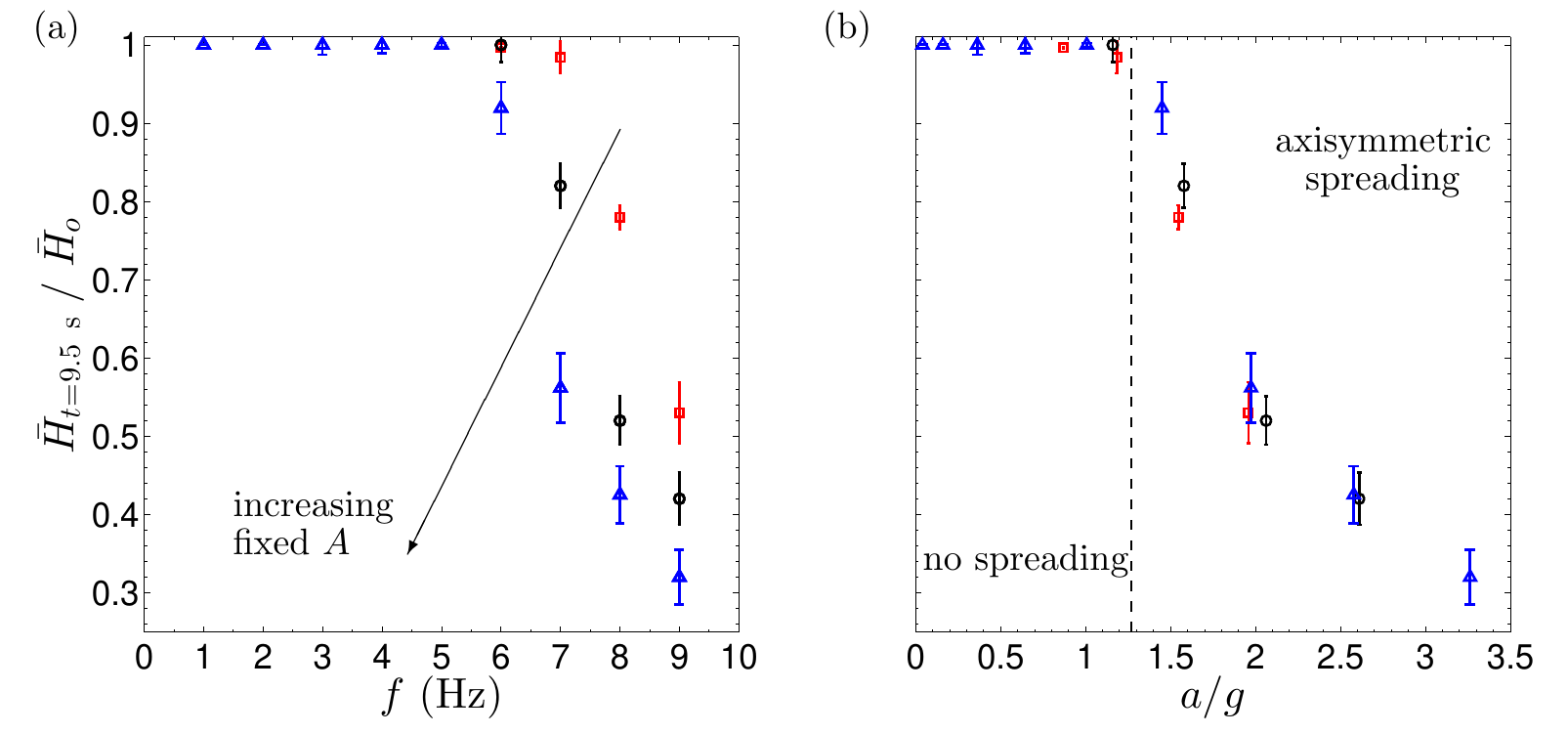}
	\caption{\small Height of the chocolate drop relative to its initial height (measured at $t=9.5$~s after imposition of the forcing) as a function of frequency (a) and acceleration (b). Each data point corresponds to the mean of at least 3 experiments and the error bar is the standard deviation. Experiments with different amplitudes of forcing are denoted by different coloured symbols: $A = 6.0$~mm (red square), $A = 8.0$~mm (black circle), $A = 10.0$~mm (blue triangle). In (b), the vertical dashed black line indicates the threshold acceleration beyond which axisymmetric spreading occurs.  This threshold was estimated by linear extrapolation of the data for $1.4 \le a/g \le 2.5$ to a unit non-dimensional height.}
	\label{fig:chocthreshold}
\end{figure} 

\subsection{Yield threshold}
\label{sec:threshold}

 \begin{figure}[!ht]
 	\includegraphics[width=0.9\textwidth]{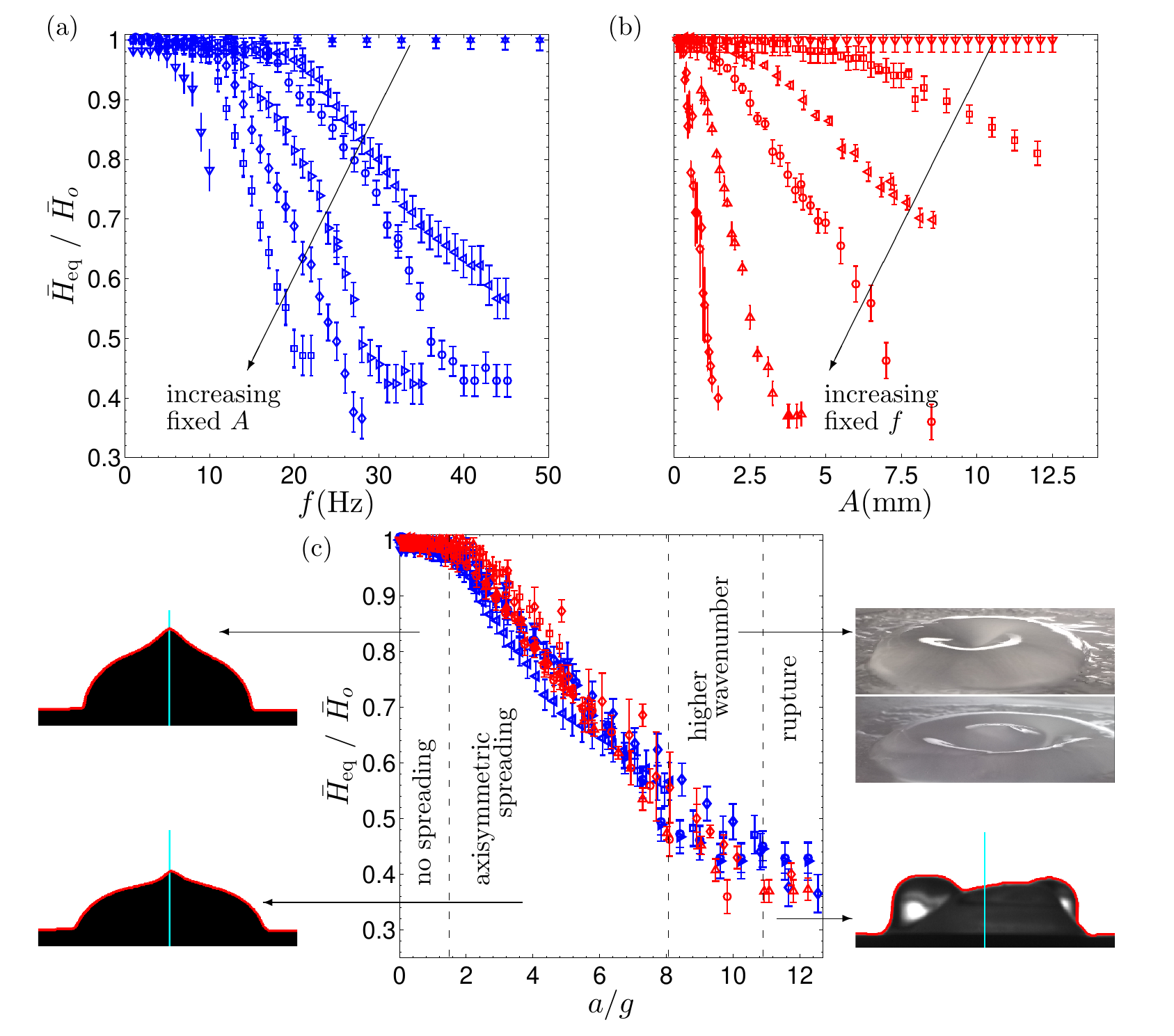}
 	\caption{\small Equilibrium height of Carbopol (2.2~g/L) drops scaled by the initial equilibrium drop height as a function of: (a) forcing frequency, (b) amplitude and (c) acceleration. Measurements were taken after interruption of the forcing imposed for $t>t_{\rm spreading}$. Each data point corresponds to the mean of at least 3 experiments and the error bars are the standard deviations. In (a), the forcing frequency is varied for fixed amplitudes (blue data): $A = 0.15$~mm (hexagon), $A = 1.0$~mm (left-pointing triangle), $A = 1.5$~mm (circle), $A = 2.5$~mm (right-pointing triangle), $A = 4$~mm (diamond) , $A = 5.5$~mm (square), $A = 12.5$~mm (downward-pointing triangle). In (b), the forcing amplitude is varied for fixed frequencies (red data): $f = 1 - 4$~Hz (downward-pointing triangle), $f = 10$~Hz (square), $f = 13$~Hz (left-pointing triangle), $f = 17$~Hz (circle), $f = 27$~Hz (upward-pointing triangle), $f = 45$~Hz (diamond). In (c), different regimes are illustrated by side-view snapshots of the Carbopol drop. The onset of spreading $a_c(Carb)/g = 1.59 \pm 0.2$ is calculated by extrapolating the data in the axisymmetric spreading region to a scaled drop height of unity. A transition to higher wavenumber oscillations occur for $a/g = 8.0 \pm 0.2$, and non-axisymmetric spreading and rupture of the drop sets in for $a/g = 10.9 \pm 0.3$.}
 	\label{fig:Carbopolthreshold}
 \end{figure} 
 
In order to determine the forcing threshold beyond which chocolate and Carbopol (2.2~g/L) drops spread, we performed experiments for a range of frequencies and amplitudes of forcing. For chocolate we measured the height of the drop at $t=9.5$~s after imposition of the forcing, by which time oscillations had decayed and most of the spreading had occurred. For Carbopol we measured the final height of the drop by stopping the forcing for $t > t_{\rm spreading}$, i.e., once the drop had spread to reach a new oscillatory state of constant mean height; see \S \ref{sec:dropspreading}. 

Results for chocolate are shown as a function of frequency in figure \ref{fig:chocthreshold}(a). Different coloured symbols are used to indicate the three datasets with different amplitudes of forcing, which are displaced from each other along the frequency axis. Figure \ref{fig:chocthreshold}(b) shows that the data collapses onto a master curve when plotted as a function of the forcing acceleration. For sufficiently small accelerations, the drop does not exhibit measurable spreading. Once the forcing acceleration exceeds a threshold value, $\bar{H}(t=9.5\;{\rm s})/\bar{H}_0$ decreases steeply with the acceleration. The threshold value for the onset of spreading, indicated with a vertical dashed line, was determined by linearly extrapolating the data for $1.4 \le a/g \le 2.5$ to yield a value of $a_{\rm c(choc)}/g =1.28 \pm 0.1$.

Similar data is reported for Carbopol in figure \ref{fig:Carbopolthreshold}. The equilibrium height $\bar{H}_\mathrm{eq}/\bar{H}_0$ is shown as a function of frequency (a) and amplitude (b) of forcing, corresponding to series of experiments performed at constant amplitude by varying the frequency, and at constant frequency by varying the amplitude, respectively. As for the chocolate experiments, individual data sets are displaced with respect to each other along the horizontal axis. Figure \ref{fig:Carbopolthreshold}(c) shows that the data approximately collapses again onto a master curve when plotted as a function of the forcing acceleration. As before, the threshold forcing acceleration for the onset of spreading was determined by linearly extrapolating the data for accelerations in the range $1.6 \le a/g \le 8$, yielding $a_{\rm c(Carb)}/g= 1.59 \pm 0.2$. 

Beyond this acceleration, the drop rapidly spread to a new equilibrium shape about which it continued to oscillate. In fact the experiments were performed by incrementally increasing either frequency or amplitude of forcing and recording the new equilibrium at each step by temporarily interrupting the forcing. For accelerations in the range $1.6 \le a/g \le 8.0$, the data points suggest a continuum of decreasing equilibrium heights corresponding to axisymmetric drops without spatial features.
For $8.0 \pm 0.2 \le a/g \le 10.9 \pm 0.3$, the drop oscillations remained axisymmetric but gained additional spatial features by transitioning to higher spatial modes. Finally, for $a/g \ge 10.9 \pm 0.3$, the oscillating drop lost its axisymmetry and exhibited localised rupture; see images in figure \ref{fig:Carbopolthreshold}(c). We did not investigate these regimes in further detail.

\begin{figure}[!t]
	\includegraphics[width=0.5\textwidth]{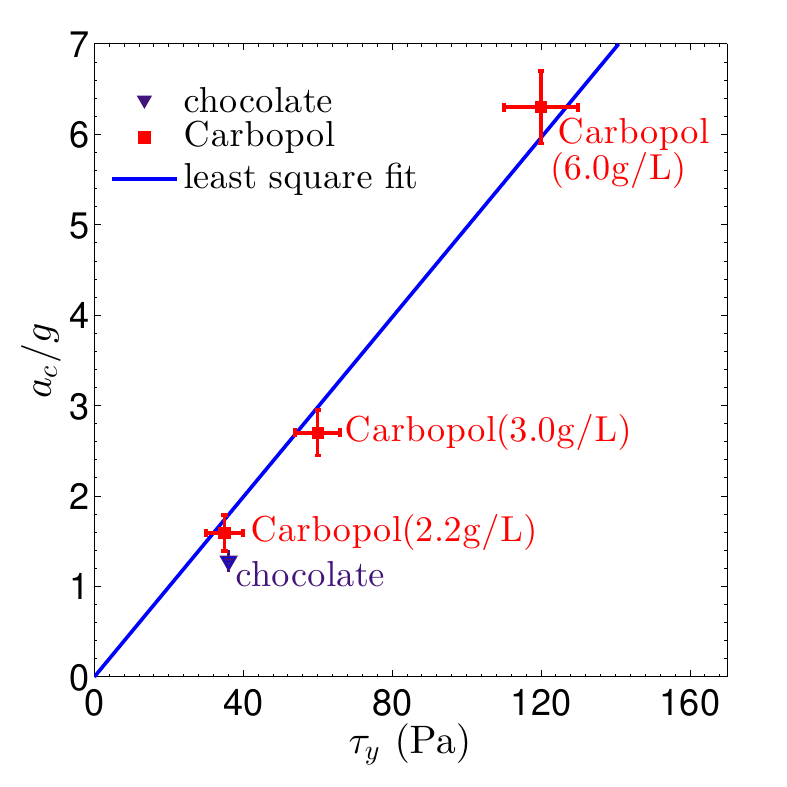}
	\caption{\small Critical forcing acceleration required to spread drops of different yield-stress fluids (see table \ref{Table1}) as a function of their yield stress. The solid line indicates a least-square linear fit to the data of the form $a_\mathrm{c}/g \propto \sigma_\mathrm{y}$.}
	\label{fig:ysa}
\end{figure}

A summary of all chocolate and Carbopol experiments is provided in figure \ref{fig:ysa} where the critical forcing acceleration is plotted as a function of yield stress (see table \ref{Table1}). We note that Carbopol (2.2~g/L) and chocolate, which have very similar values of the yield stress, also exhibit similar values of the critical acceleration. We find that the critical acceleration increases approximately proportionally to the yield stress as shown by the solid blue line, which is a one-parameter linear fit to the data. 
This confirms that the yield stress uniquely determines the onset of spreading in our experiments. However, the existence of a critical acceleration for the onset of spreading is in itself significant. If prior to the imposition of oscillatory forcing, the drop had reached its equilibrium by slumping due to gravity forces alone, we would expect the maximum stress in the drop to have relaxed to $\tau_y$ at equilibrium. Hence, any additional acceleration imposed during the oscillatory forcing cycle would be sufficient to fluidise the drop. Observation of a critical acceleration suggests that the initial stress distribution within the drop must be significantly below yield because an additional acceleration greater than that of gravity has to be imposed in order for the drop to spread. This implies that the constant of proportionality between critical acceleration and yield stress shown in figure \ref{fig:ysa} is set by the process by which the drop is generated. The vibrated-drop experiment can be therefore be used to determine the yield stress of other materials based on measured critical acceleration values provided that the same drop deposition process is retained and that the initial drop shape are similar. 

\subsection{Dependence of Carbopol spreading on the history of forcing}
\label{sec:timehistory}

\begin{figure}[!ht]
	\includegraphics[]{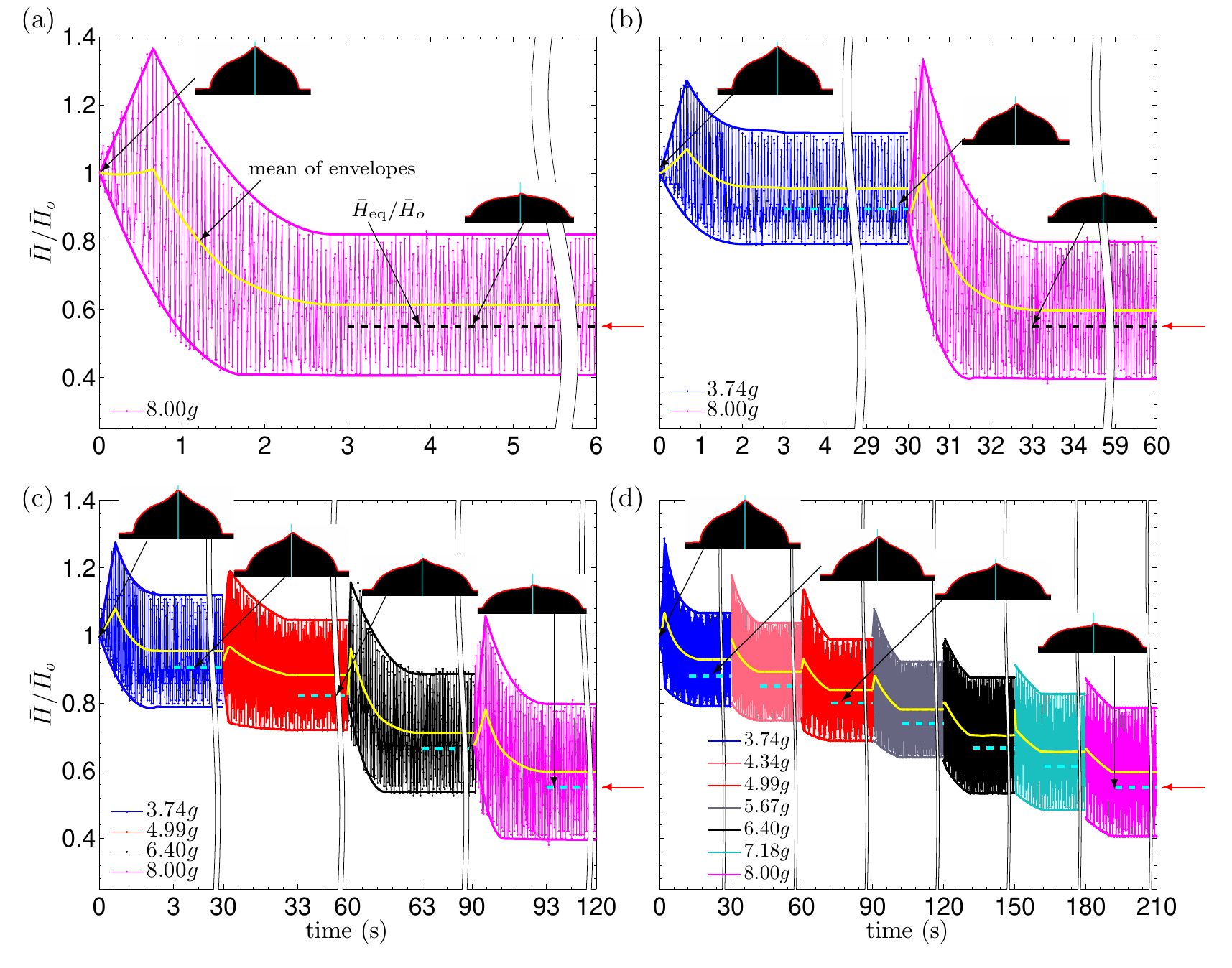}
	\caption{\small Influence of the forcing history on the spreading of a Carbopol drop (2.2~g/L), starting from the sessile configuration shown in figure \ref{fig:sessile}. The final forcing is $a=8.00g$ in each panel. This forcing is applied directly in (a), and following 1, 3 and 6 increments in (b--d). After each period of spreading, the forcing was interrupted to measure the new equilibrium shape of the drop (inset images). The envelopes of the drop oscillations are shown with solid lines of the same colour as the data, while the mean of the envelopes is shown with a yellow solid line. The successive equilibrium heights of the drop are indicated with dashed lines. The red arrows highlight that all forcing protocols result in the same final equilibrium drop heights (and profiles).}
	\label{fig:four}
\end{figure} 

We showed in \S \ref{sec:dropspreading} that Carbopol drops spread when the value of the forcing acceleration exceeds a critical threshold. We now turn to investigating the effect of forcing history upon the final equilibrium shape adopted by the drop post-spreading.

Figure \ref{fig:four} shows four spreading experiments on a Carbopol drop (2.2~g/L) starting from the sessile shape of height $\bar{H}_0$ shown in \S \ref{slump}. In (a) the drop was subjected to a forcing acceleration $a=8.0g$ for the entire duration of the experiment. In (b--d), this acceleration was reached via 1, 3 and 6 acceleration increments, respectively. The forcing was temporarily interrupted after each increment  in order to record the new equilibrium height $\bar{H}_\mathrm{eq}$ of the drop. Despite these different experimental procedures, the value of $\bar{H}_\mathrm{eq}/\bar{H_0}$ following interruption of the forcing at $a=8.0g$ is the same in all experiments, as indicated by the dashed horizontal lines on the pink experimental data (highlighted by red arrows). This indicates that spreading of the drop is not sensitive to the forcing history. 
It suggests that upon exceeding the critical acceleration, the Carbopol drop rapidly reaches a new equilibrium shape about which it oscillates periodically while the stress remains below the yield stress. Because this new drop equilibrium is at its yield threshold, any subsequent increase in forcing acceleration causes the drop to spread further to another flatter equilibrium shape.  Further experiments with forcing accelerations in the range $1.70g \le a \le 9.68g$ and between 1 and 14 acceleration increments (not shown) confirmed that the final equilibrium height of the drop is indeed independent of the time-history of forcing to within experimental resolution provided that all experiments start from the same initial sessile drop shape.


\begin{figure}[!ht]
	\includegraphics[]{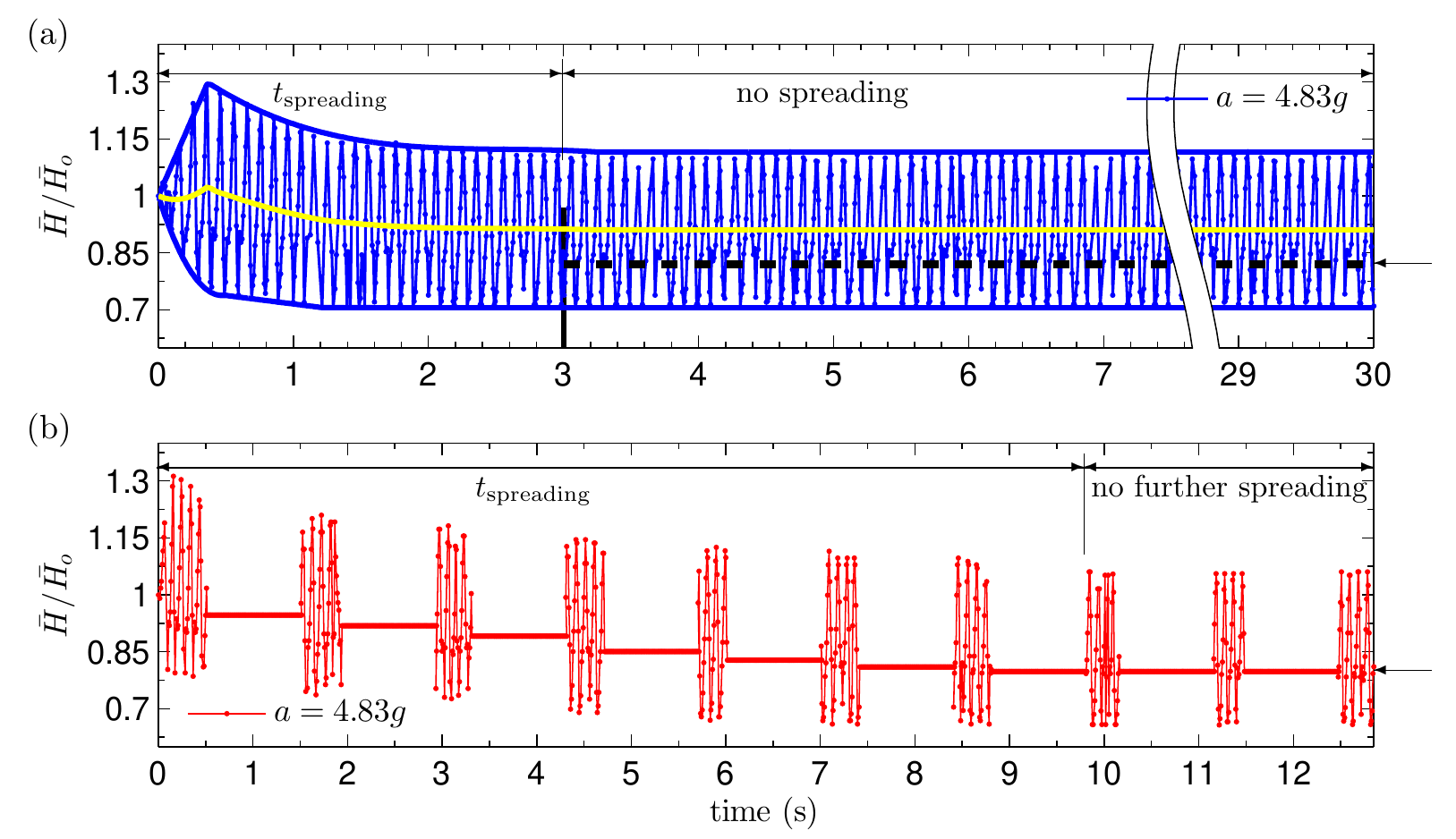}
	\caption{\small Spreading of a Carbopol drop (2.2~g/L) for $a=4.83g$ ($f = 10$~Hz and $A = 12$~mm) when the forcing was applied continuously (a) and intermittently (b). In (b), the duration of the intermittent bursts of oscillatory forcing are less than $t_\mathrm{spreading}\simeq 3.2$~s required for the droplet to reach an oscillatory state about a new equilibrium. Each time the forcing is interrupted, the drop settles into a new equilibrium state. The total duration of the intermittent oscillations during which spreading occurs in (b) is approximately equal to the continuous spreading time in (a), with small effects of start-up and shut-down of the forcing visible in (b).}
	\label{fig:burst}
\end{figure} 

We also examined the effect of applying forcing in short, successive bursts of 5 periods of oscillation of the substrate. This meant that the forcing was interrupted before the drop had spread to a new equilibrium shape. The comparison in figure \ref{fig:burst} between continuous forcing (a) and short bursts (b) at $a=4.83g$ shows that spreading occurs either continuously or intermittently, but that the drop reaches the same long-term equilibrium (highlighted by the black arrows) about which the drop can oscillate. Remarkably, the cumulative time required to spread to the new equilibrium while the forcing is applied is approximately the same ($\simeq 3$~s)  in both cases, although in figure \ref{fig:burst}(b) the ramp-up and ramp-down of the forcing result in small alterations of the forcing signal.

\begin{figure}[!ht]
	\includegraphics[]{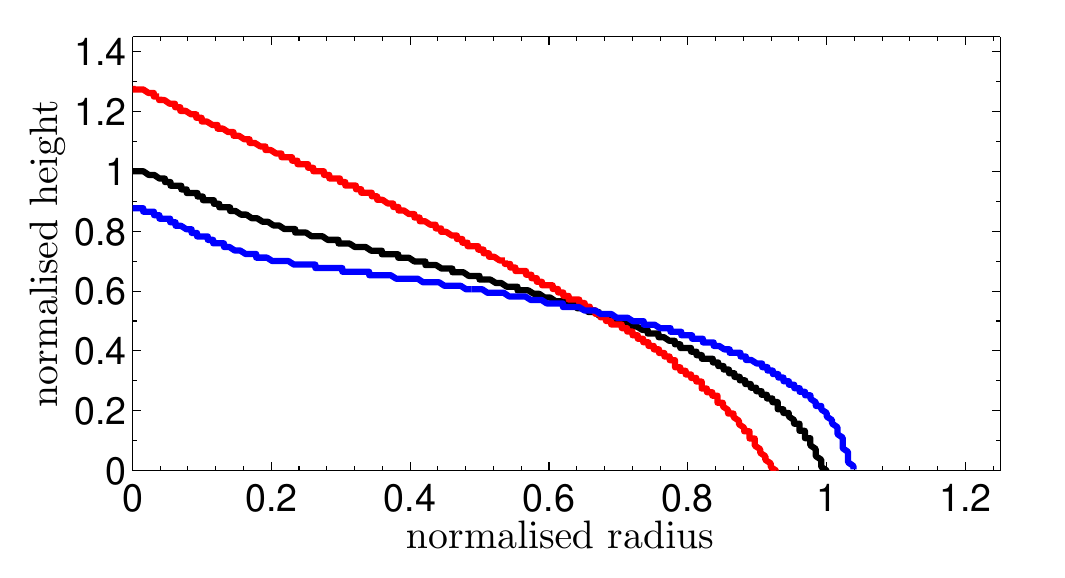}
	\caption{\small Comparison between the drop shape upon interruption of the forcing and its final equilibrium shape after relaxation. The height of the drop normalised by its equilibrium height is plotted as a function of the normalised radius. The blue and red lines show two instantaneous drop shapes at the end of the substrate motion. They both relax to the same final equilibrium shape (black).}
	\label{fig:stop}
\end{figure}

In a final set of experiments, we investigated the influence of the position at which the substrate came to rest on the equilibrium shape of the drop reached in figure \ref{fig:burst}. We imposed approximately two periods of oscillation of the substrate (with additional ramp-up and down) at $a=4.83g$, starting with the substrate either in its neutral position, its bottom position or its top position. We then ensured that the substrate came to rest in each of these three positions. Figure \ref{fig:stop} shows that the ultimate equilibrium shape of the drop reached after interruption of the forcing was the same irrespective of the forcing protocol. 

\subsection{Viscoelastic oscillations of the Carbopol drop}
\label{sec:viscoelastic}

\begin{figure}[!ht]
	\includegraphics[width=\textwidth]{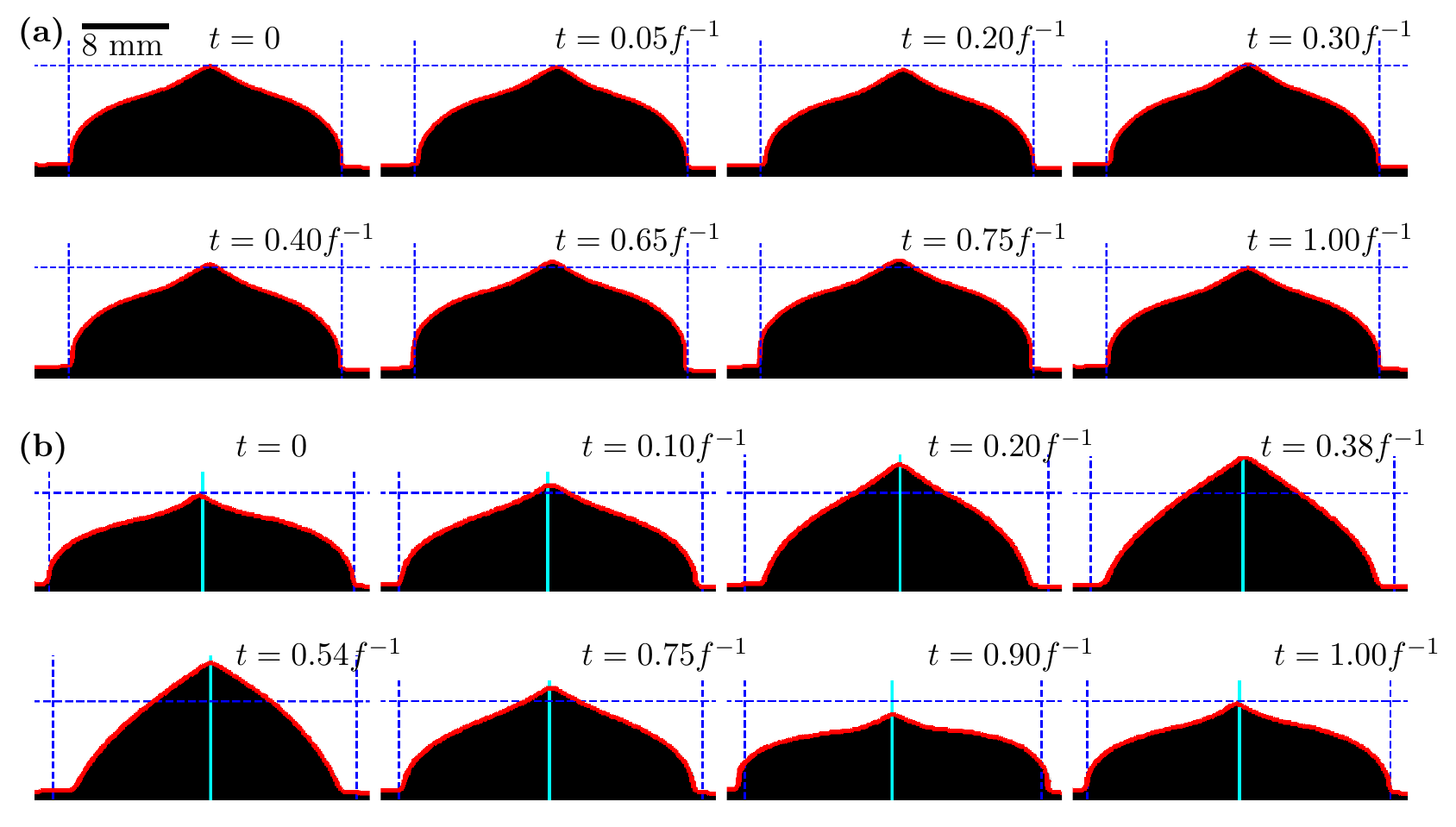}
	\caption{\small Time-periodic deformation of a Carbopol drop (2.2~g/L) under vertical vibration during one period of oscillation $1/f$ for (a) $a= 0.86g < a_c$ ($f=5$~Hz and $A=8.5$~mm) and (b) $a = 7.11g> a_c$ ($f = 14$~Hz, and $A = 9$~mm), after the drop has reached a state of constant mean height following spreading. The processed images show the drop interface and the platform edge (red line). The dashed blue lines, which indicate the height and diameter of the initial drop at $t=0$, highlight the compression and extension of the drop over the period of oscillation. }
	\label{fig:harmonic}
\end{figure} 


We have established that upon exceeding a critical forcing acceleration, Carbopol drops fluidise and rapidly spread until they reach a new equilibrium drop shape about which they continue to oscillate. These observations suggest that the change in equilibrium drop shape due to spreading results in a reduction of the stress within the drop to below the yield stress, so that Carbopol recovers the behaviour of a viscoelastic solid. In this section, we investigate the viscoelastic oscillations of the drop about the different equilibrium states that are reached as the oscillatory forcing is increased. Figure \ref{fig:harmonic} shows typical Carbopol (2.2~g/L) drop shapes over one period of the oscillation for $a<a_c$ (a) and $a>a_c$ following the initial spreading phase (b). The maximum variation in the height of the drop over the period of the oscillation increases significantly with the imposed acceleration. Whereas the height oscillation is barely visible for $a=0.86g<a_c$ (figure \ref{fig:harmonic}(a)), when $a=7.11g> a_c$ (figure \ref{fig:harmonic}(b)) the drop extends significantly above its equilibrium height (dashed blue horizontal line) and reduces its diameter below its equilibrium value (e.g., at $t=0.38f^{-1}$). Conversely, during compression the height change is smaller but an increase in the diameter is still clearly visible ($t=0.90f^{-1}$). In addition, the phase of the drop's response differs 
for the two values of oscillatory forcing in that the drop extends vertically with the rising platform for $a>a_c$ while it compresses with the rising platform for $a<a_c$. 

\begin{figure}[!ht]
	\includegraphics[]{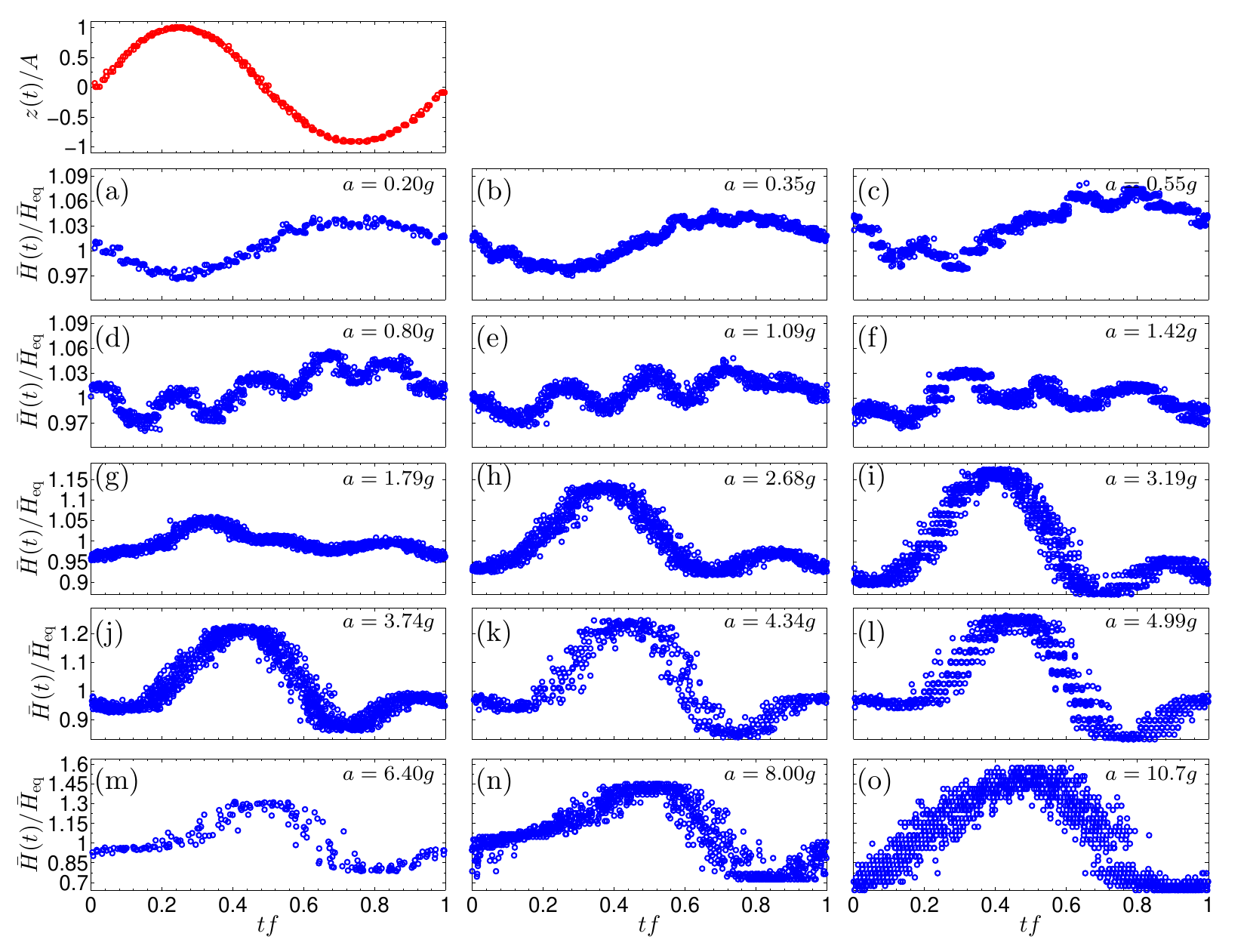}
	\caption{Time series of the height of a Carbopol drop (2.2~g/L) relative to its equilibrium height (blue) for increasing values of forcing acceleration. Each time series represents multiple cycles of oscillation over one period $f^{-1}$ of the substrate oscillation which is shown in red in the top panel (in terms of the normalised substrate displacement as a function of dimensionless time $tf$). The forcing amplitude is $A=5.5$~mm. The frequencies are (a) $f=3$~Hz, (b) $f=4$~Hz, (c) $f=5$~Hz, (d) $f=6$~Hz, (e) $f=7$~Hz, (f) $f=8$~Hz, (g) $f=9$~Hz, (h) $f=11$~Hz, (i) $f=12$~Hz, (j) $f=13$~Hz, (k) $f=14$~Hz, (l) $f=15$~Hz, (m) $f=17$~Hz, (n) $f=19$~Hz and (o) $f=22$~Hz. Note the increase in vertical axis range in successive rows.}
	\label{fig:timeseries}
\end{figure} 

\begin{figure}[!ht]
	\includegraphics[]{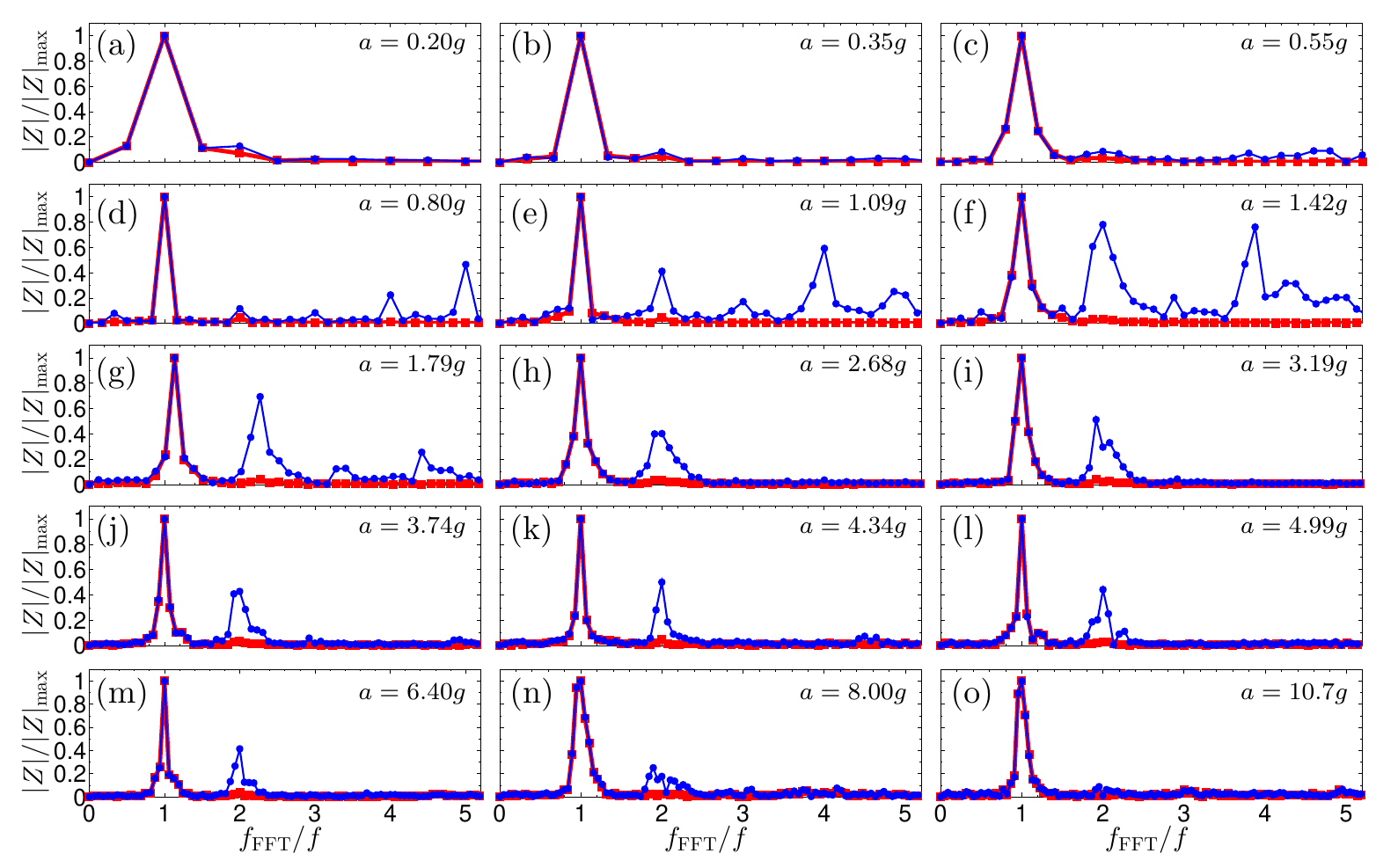}
	\caption{Comparison between the normalised Fast Fourier Transform of the time series of drop height shown in figure \ref{fig:timeseries} (blue) and the vertical displacement of the platform (red). Forcing parameters are the same as in figure \ref{fig:timeseries}.}
	\label{fig:fft}
\end{figure} 

\begin{figure}[!ht]
	\includegraphics[]{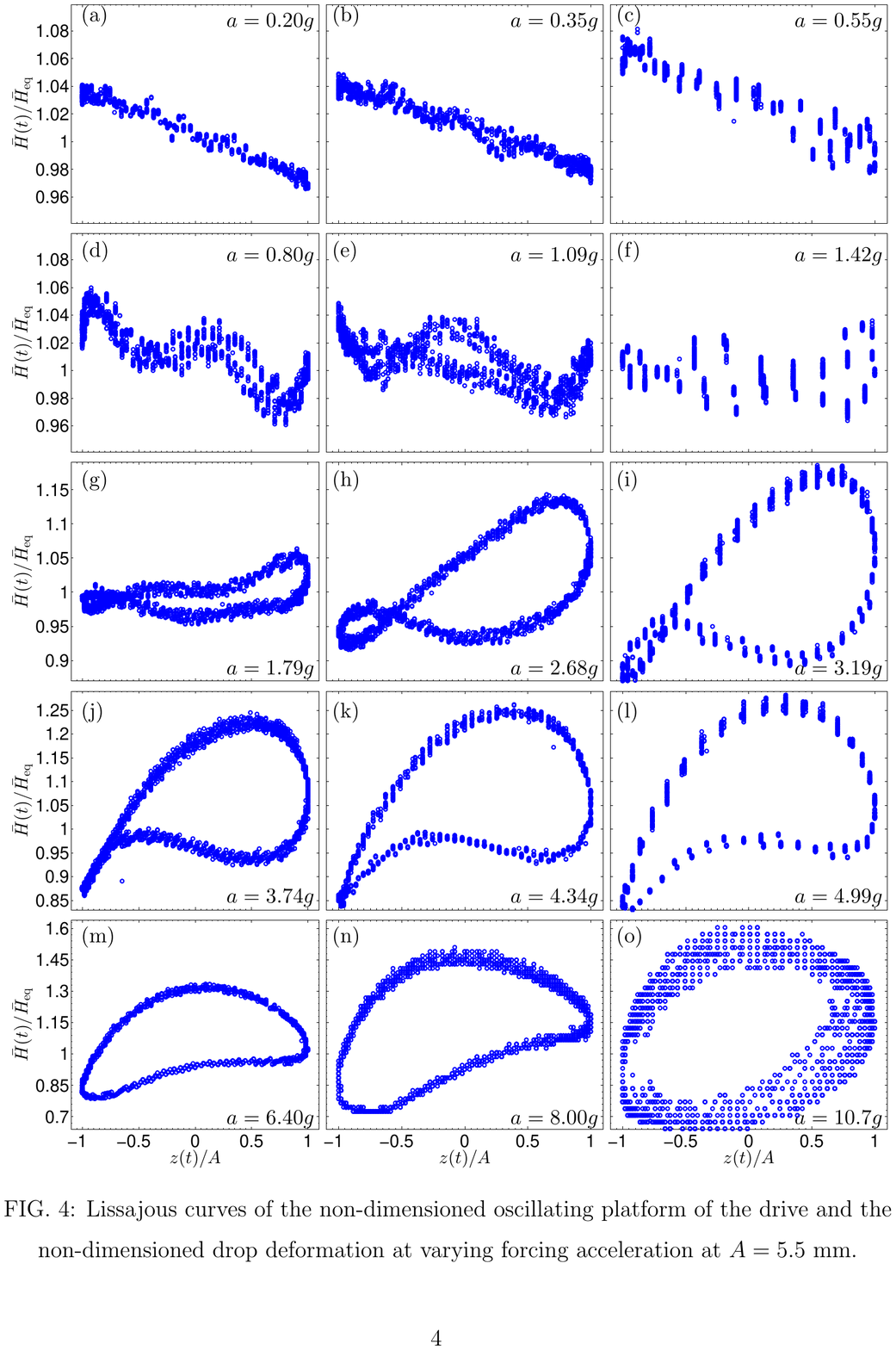}
	\caption{Lissajous curves of the data presented in figure \ref{fig:timeseries} obtained by plotting the height of the drop relative to its equilibrium height as a function of the vertical displacement of the platform. Forcing parameters are the same as in figure \ref{fig:timeseries}.}
	\label{fig:Lissajous}
\end{figure} 

Figure \ref{fig:timeseries} shows a sequence of time series of the drop height (blue) for increasing values of the forcing acceleration. Multiple cycles are plotted over a single period of the sinusoidal forcing, which is shown in red in the top panel.
For the lowest accelerations (a, b), the oscillations of the viscoelastic solid drop are approximately sinusoidal and lag approximately $180^\circ$ behind the position of the substrate. 
For $a<a_c$,  the peak-to-peak amplitude of the drop oscillations appears to vary only weakly with forcing acceleration.
However, figures \ref{fig:timeseries}(c--f) indicate a weakening of the fundamental mode of frequency $f$, and the appearance of higher frequency modes in the signal as the forcing acceleration is increased towards the yield threshold. This change in the drop response is quantified in figure \ref{fig:fft}: the normalised Fast Fourier Transforms (FFT) of the time series shown in figure \ref{fig:timeseries} reveal harmonics up to $5f$. Figure \ref{fig:fft} indicates that higher harmonics first appear in (c) and their contribution increases considerably in (d--f) so that the amplitude of the first harmonic of frequency $2f$ increases and approaches that of the fundamental mode of frequency $f$ as the forcing acceleration reaches $a=1.42g < a_c$. Although the harmonic content of the signal progressively decreases as $a$ increases above $a_c$, the first harmonic persists up to large forcing accelerations (g--n) and becomes indistinguishable from the signal noise only for $a=10.7g$ (o). The rich harmonic content of the drop response near the yield threshold suggests significant nonlinear viscoelastic effects in this region. 

Returning to the time-series of figure \ref{fig:timeseries}, the gradual reduction in the amplitude of the fundamental mode up to $a=1.42g$ (f) is reversed in (g) (note the change in the vertical axis range) where it begins to strengthen again but with a phase-shift of approximately $100^\circ$ compared with (a--f). 
A pictorial illustration of this evolution is provided by the Lissajous curves in figure \ref{fig:Lissajous}, where the drop height relative to its equilibrium value is plotted as a function of the vertical displacement of the substrate. In (a--c), the maximum and minimum drop heights occur for the minimum and maximum vertical substrate positions, respectively. The limit cycle (which appears as a line in (a) due to the small amplitude of the drop oscillations) then rotates anti-clockwise with increasing forcing acceleration towards an approximately horizontal structure (d--g). This rotation continues in (h--l) so that the maximum drop extension and compression occur near the maximum and minimum substrate heights, respectively. The strong second harmonic component of the signal means that the cycle is a distorted figure of eight (g--j), which evolves into a simple cycle as the second and higher harmonics subside. In (m,n), the maximum compression of the drop still occurs at the minimum vertical substrate position but the maximum extension occurs closer to the neutral substrate position. In (o) the cycle has reshaped so that the maximum compression of the drop occurs increasingly close to the neutral position of the substrate and the response of the drop to the forcing is now close to sinusoidal again. 

\begin{figure}[!ht]
	\includegraphics[]{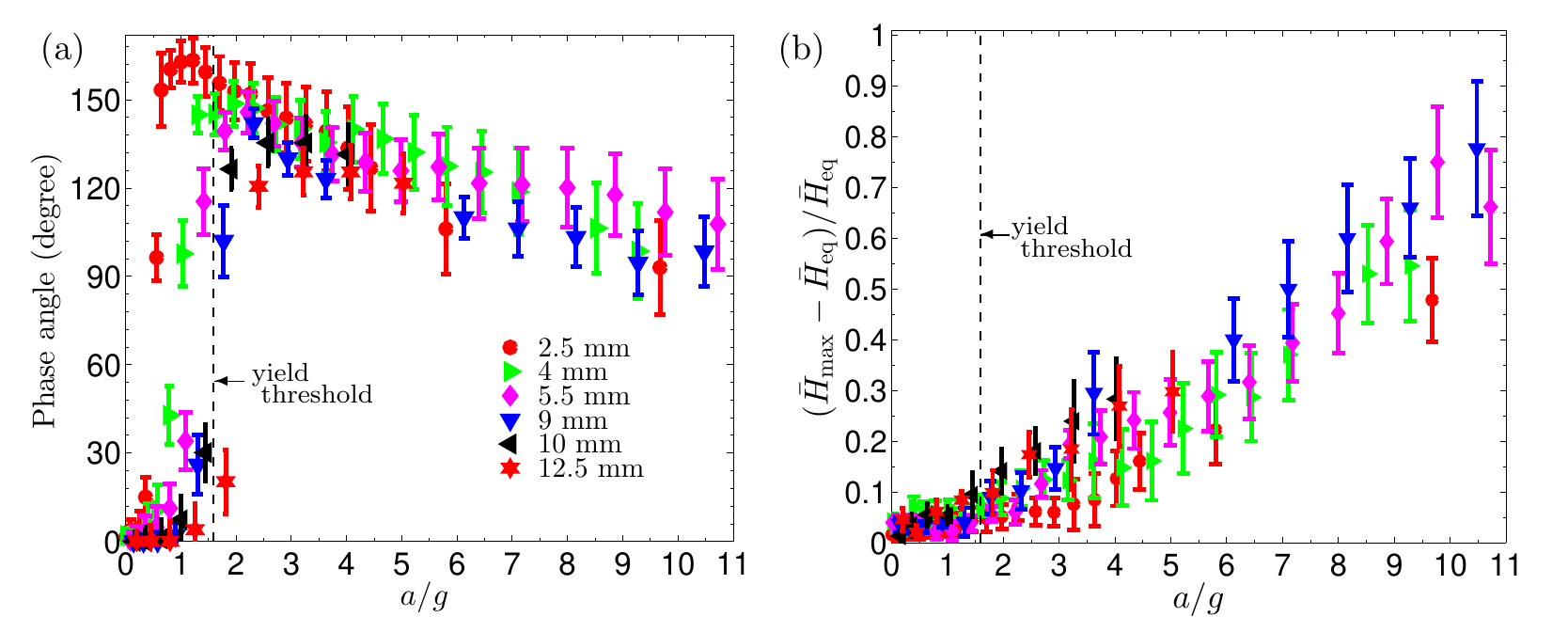}
	\caption{Dependence on forcing acceleration of (a) the phase difference between drop oscillations and vertical substrate acceleration, (b) the maximum extension of the drop. Each symbol corresponds to a fixed amplitude of oscillation, so that increasing values of forcing acceleration are achieved by increasing the frequency of forcing. The error bars in (b) indicate the standard deviation of at least three experiments.}
	\label{fig:phaseamp}
\end{figure} 

We summarise the dependence of the viscoelastic oscillations of Carbopol drops (2.2~g/L) on the forcing acceleration in figure \ref{fig:phaseamp}. Figure \ref{fig:phaseamp}(a) shows the phase lag of the fundamental harmonic of the oscillatory drop response relative to the substrate acceleration, which was determined from the FFT data. The phase angle increases steeply from values near $0^\circ$ to nearly $180^\circ$ on the approach to the yield threshold as previously observed in figures \ref{fig:timeseries} and \ref{fig:Lissajous}. 

Figure \ref{fig:phaseamp}(b) shows that the maximum extension of the drop increases monotonically with forcing acceleration. The maximum compression of the drop follows a similar relation (not shown). The maximum extension (and the smaller compression) provide measures of the maximum strain in the drop, by accounting for the decrease in equilibrium height of the drop which occurs through spreading as the forcing acceleration is increased.

\section{Discussion and conclusion}
\label{disc}

We have studied experimentally the response to sinusoidal vertical displacement of drops of molten chocolate and Carbopol initially at rest on a thin layer of the same fluid. These shear-thinning yield-stress fluids have very different pre-yield elastic moduli, up to a factor 100 larger for chocolate than for Carbopol, because of their different mesostructures. Carbopol is also more strongly shear-thinning with a steeper decrease of its viscosity to lower values than chocolate. We adjusted the Carbopol concentration so that its yield stress matched that of chocolate and found that drops of both materials first spread for approximately the same value of forcing acceleration $a_c$. For $a<a_c$, the chocolate drop is at rest in the frame of reference of the oscillating substrate while the Carbopol drop undergoes viscoelastic stretching and compression periodically with the forcing. For $a\ge a_c$,
rapid initial spreading of the chocolate drop gives way to long-term slower motion, whereas in Carbopol, the drop rapidly relaxes its stress by spreading to a new equilibrium shape of larger footprint, which continues to undergo large-amplitude viscoelastic stretching and compression. Similar viscoelastic oscillations are also observed transiently in the chocolate drop for sufficiently large forcing but they become weaker as the drop spreads. This reduction in overall strain is consistent with an increase of the elastic modulus, which is known to occur in chocolate over successive fluidisation cycles through mesoscopic reorganisation of the material \cite{bergemann:2018b}. The strong viscoelastic effects observed in Carbopol result in a striking reduction in spreading: for $a=3.25g$, the total height reduction of the Carbopol drop is 12\% compared with 70\% for the chocolate drop after 8~s, which then only spreads weakly at larger times; see figure~\ref{fig:dynamics_chocolate_carbopol_above}. 

The existence of a forcing threshold for the onset of spreading in our experiment indicates that the initial drop shape at equilibrium under gravity (reached after deposition from the syringe) has a stress distribution below the yield stress. We found that the threshold for the onset of spreading is uniquely determined by a critical forcing acceleration $A (2 \pi f)^2$ and is proportional to the yield stress. In practice, this means that our experimental setup can be used to measure the yield stress of materials provided that prior calibration of the rate of increase of forcing acceleration with yield stress is performed with a fluid of known yield stress.  Our threshold criterion differs from that of experiments at much larger forcing \cite{shiba:2007}, where yield phenomena occur above a critical velocity of forcing, i.e. when the stress $\rho (2\pi f A)^2$ exerted by the substrate due to inertia of the material exceeds the yield stress. 

\begin{figure}[!ht]
	\includegraphics[]{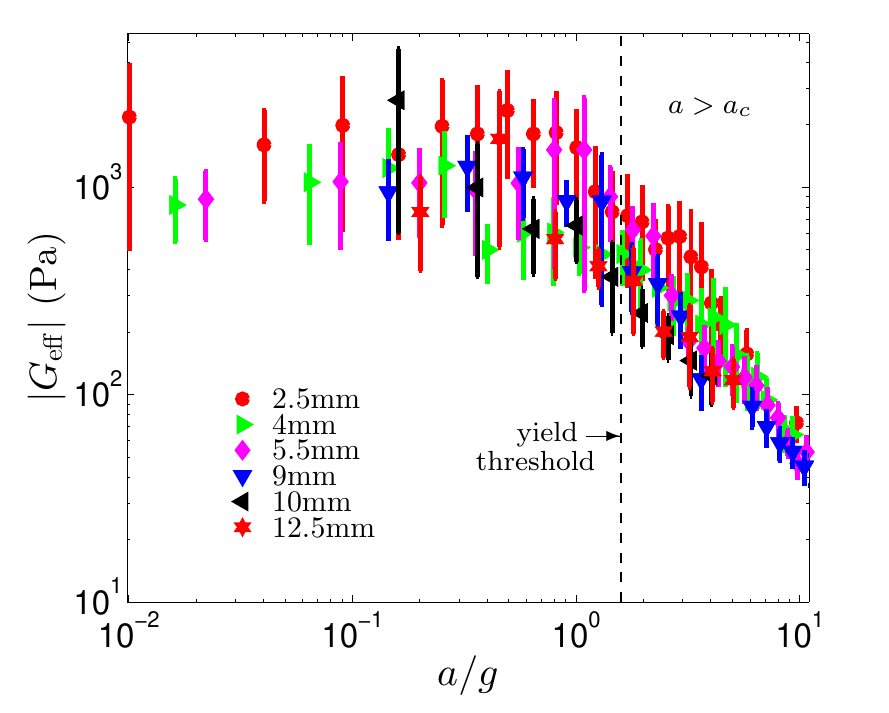}
	\caption{Estimate of the variation of the absolute value of the effective complex modulus $|G_\mathrm{eff}|$ of Carbopol as a function of forcing acceleration.}
	\label{fig:Gvibrated}
\end{figure} 

\begin{figure}[!ht]
	\includegraphics[]{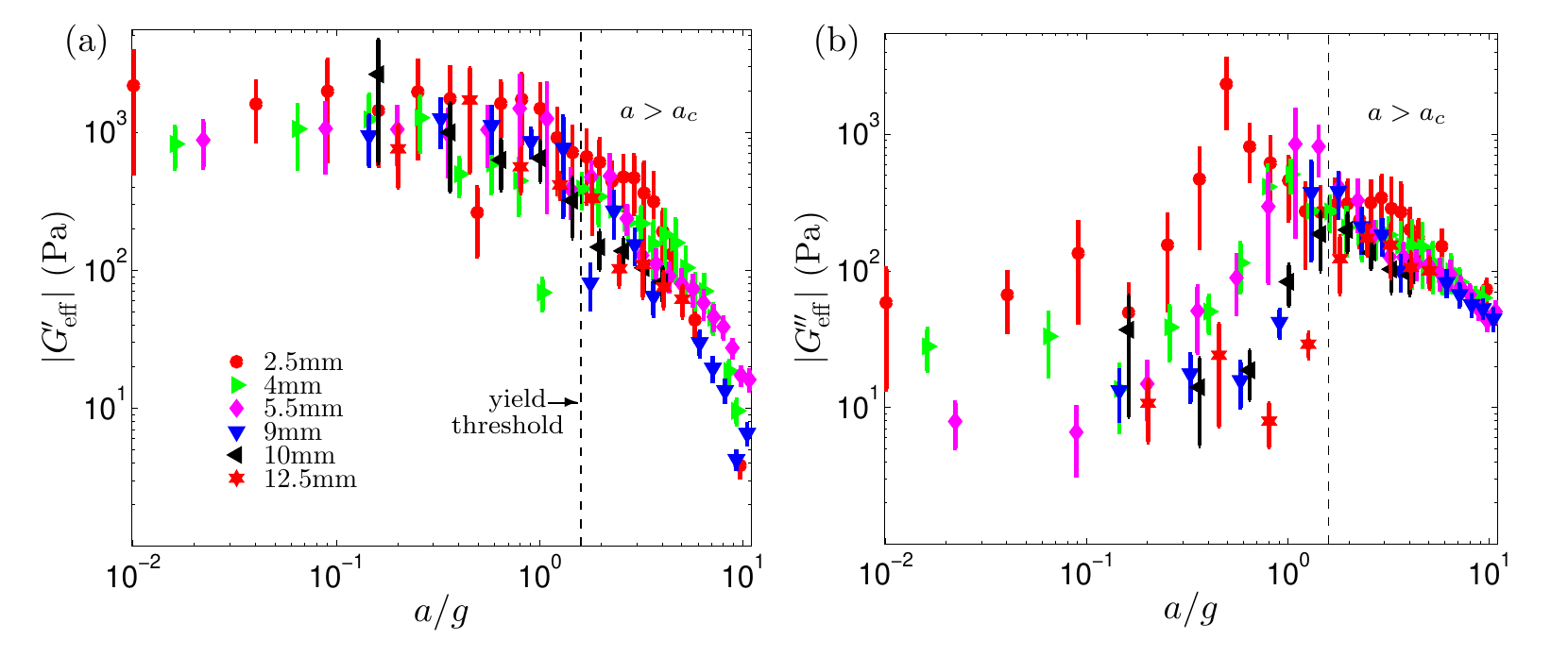}
	\caption{Estimate of the variation as a function of forcing acceleration of the effective storage (a) and loss moduli  (b), which are the real and imaginary parts of the effective complex modulus of Carbopol and are calculated using the data in figure \ref{fig:phaseamp}.}
	\label{fig:GG}
\end{figure} 

Remarkably, we found that the spreading of Carbopol drops did not depend on the history of forcing. We were able to interrupt and resume forcing as well as increase the forcing acceleration with multiple increments without affecting the final equilibrium drop shape. If the forcing was discontinued before a new equilibrium drop shape had been reached, spreading towards this equilibrium state continued when reapplying the forcing. This confirms that spreading continues until the stress within the drop has reduced to the yield stress, which is indicated by a new equilibrium drop shape undergoing large amplitude viscoelastic oscillations. 

Hence, the Carbopol drop (and presumably also the chocolate drop) automatically adjusts its shape to remain at the yield stress for increasing values of the forcing acceleration $a\ge a_c$. Our experiment is therefore uniquely taylored to investigate the rheology of the material at its yield stress. Although our experimental evidence suggests that after each increase in forcing acceleration the material returns to a viscoelastic solid state following short transients, we cannot exclude the existence of recirculating viscoplastic flows within the drop. \citet{shiba:2007} observed free-standing convection rolls in vibrated drops of a different elastoviscoplastic gel in addition to viscoelastic oscillations, but the convective flow was associated with time scales much longer than the period of the forcing. We also find a complex nonlinear viscoelastic response of the drop to the oscillatory forcing near $a_c$: a steep increase in the phase angle of the fundamental oscillation mode and oscillations at higher harmonic frequencies. 

Increasing the forcing acceleration increases the amplitude of the drop oscillations about the equilibrium state and thus the strain in the drop, without exceeding the yield stress. We use the data of figure \ref{fig:phaseamp}(b) to estimate the absolute value of the effective complex modulus of the Carbopol microgel as 
$$|G_\mathrm{eff}|= \frac{\tau_y} {(\bar{H}_\mathrm{max}-\bar{H}_\mathrm{eq})/\bar{H}_\mathrm{eq}},$$
 which is plotted as a function of forcing acceleration in figure \ref{fig:Gvibrated}. We also decompose the effective complex modulus into the storage and loss moduli $G'_\mathrm{eff}$ and $G''_\mathrm{eff}$ in figure \ref{fig:GG} using the phase angle shown in figure \ref{fig:phaseamp}(a). The most striking feature of these plots is the monotonic decrease within our parameter range of $|G_\mathrm{eff}|$ and $G'_\mathrm{eff}$.  Remarkably, the values of $G'_\mathrm{eff}$ and $G''_\mathrm{eff}$ follow similar qualitative trends to the storage and loss moduli of Carbopol microgels as functions of either imposed strain or stress, measured by oscillatory-shear rheometry \cite{fernandes:2017,varges:2019,digiuseppe:2015}. Although the values obtained for $a<a_c$ overestimate $|G_\mathrm{eff}|$ because the stress within the drop is less than the yield stress, $|G_\mathrm{eff}|$ is of the correct order of magnitude of hundreds of Pa (see table \ref{Table1}). Hence, our vibrated drop offers a route to simple measurements of the nonlinear viscoelastic properties of the material at the yield stress. We also find that the strain in the drop eventually reaches a maximum with increasing forcing acceleration for $a \ge 10.9g$ where the drop ruptures.
 We therefore conclude that our vibrated-drop setup offers unique features to help deepen insight into the rheology of elastoviscoplastic fluids at their yield threshold 
 and could act as a simple rheometer for fluids undergoing large viscoelastic deformations. In order to deepen understanding of the drop behaviour, it would be particularly interesting to explore whether the experimental dynamics reported in this paper can be reproduced numerically using an appropriate constitutive model such as the Saramito \cite{saramito:2009} model extended to include nonlinear viscoelasticity.

\appendix

\section{Quasi-steady shear rheometry} \label{AppA}

We performed quasi-steady shear-rheometry measurements of the yield stress and viscosity curves of the Carbopol microgels used in our experiments by applying the methodology of \citet{bergemann:2018b}. We used a Brookfield R/S-Plus (SST) rheometer with a four-bladed vane spindle (blade height of $20\pm0.01$~mm) rotating in a stationary cylindrical glass beaker of inner radius $42.5\pm0.25$~mm, filled with
Carbopol microgel to a height of $50\pm2$~mm. We refer the interested reader to \cite{bergemann:2018b} for details of the experimental protocol and associated parameter values.

\begin{figure}[!ht]
\includegraphics[]{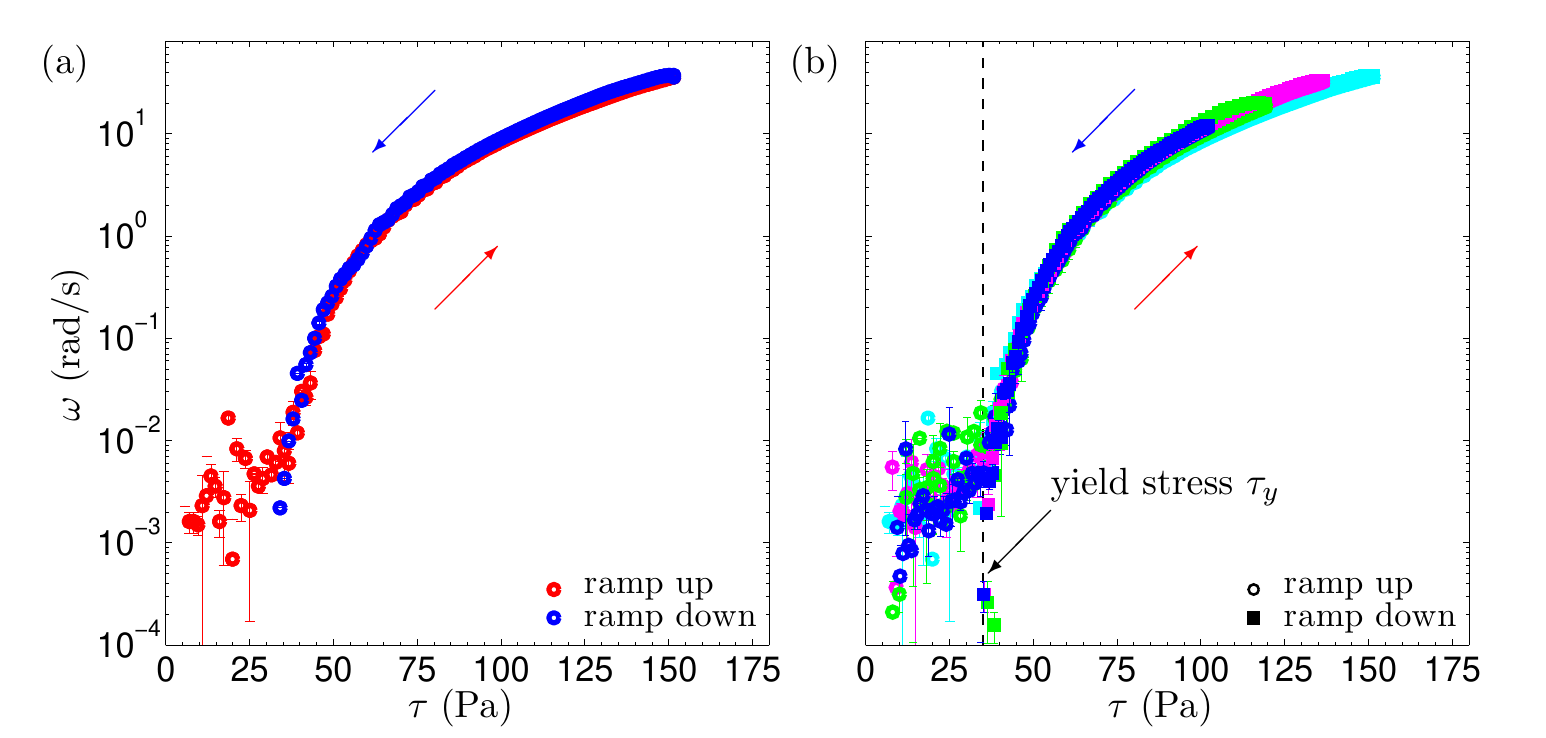}
	\caption{\small Measured angular velocity as a function of shear stress in Carbopol microgel (2.2~g/L). Shear stress was imposed periodically through linear ramp-up and down phases indicated in (a) by red and blue symbols and in (b) by circles and squares, respectively. (a) Cycle period of $\tilde{T} = 240$~s and ramping rate $\alpha=1.33$~Pa/s. (b) Ramping rates and cycle periods of $\alpha=0.83$~Pa/s, $\tilde{T} =240$~s (blue), $\alpha=1.00$~Pa/s, $\tilde{T} =240$~s (green), $\alpha=1.16$~Pa/s, $\tilde{T} =240$~s (magenta), $\alpha=2.67$~Pa/s, $\tilde{T} =120$~s (cyan).The data shown corresponds to mean values indicate of 4 repetitions of individual experiments of 5 cycles, and the error bars are the standard deviations from the mean. The vertical black line in (b) indicates the value of the dynamic yield-stress $\tau_y$ of the material.}
	\label{fig:dynamics_chocolate_carbopol_3}
\end{figure}

In order to measure yield stress, we imposed cycles of linearly increasing and decreasing torque and measured the angular velocity. The shear stress $\tau$ was readily calculated from the imposed torque and the
geometry of the spindle, with a correction for spindle end-effects applied by adding a virtual length to actual spindle length \cite{bergemann:2018b}. Figure \ref{fig:dynamics_chocolate_carbopol_3}(a) shows the measured angular velocity of the spindle as a function of the applied shear stress for the 2.2~g/L Carbopol microgel. The period of the forcing cycle was $240$~s and the maximum stress applied in each cycle was 160~Pa. The error bars correspond to standard deviations over 5 consecutive cycles of 4 repetitions of the experiments.

During the ramp-up phase (red symbols), a modest increase of the angular velocity $\omega$ for $\tau \lesssim 30$~Pa is followed by a steep and smooth increase beyond this threshold.  Large fluctuations in $\omega$ occur for $\tau \lesssim 30$~Pa, which were absent in chocolate \cite{bergemann:2018b}. 
For $\tau \lesssim 30$~Pa, the Carbopol microgel deforms as a solid, and the large variability in the rotation rate measurements can be attributed to viscoelasticity and stick-slip of the Carbopol microgel on the spindle walls \cite{ovarlez:2013,balmforth:2014,bonn:2017,birren:2019}. The ramp-down phase (blue symbols) shows a smooth decrease of $\omega$ which retraces the red curve to within 5\%. However, the blue dataset continually steepens with decreasing stress. Thus, it is associated with a minimum stress value required for maintaining flow known as the dynamic yield-stress, $\tau_y = 35 \pm 3$~Pa, which we obtained by extrapolation of the blue dataset. 

Figure \ref{fig:dynamics_chocolate_carbopol_3}(b) shows that our results are insensitive to the details of the ramping cycle. Experiments performed with different rates of change of the applied stress, $0.83 \le \alpha \le 2.67$~Pa/s, and cycle periods of either 120 and $240$~s all collapse approximately onto a master curve above the yield stress. The hysteresis between the ramp-up and ramp-down parts of the forcing cycle is on the order of the measurement error, which suggests that thixotropy is negligible within our parameter range.
The dynamic yield stress estimated from all the experiments performed with the 2.2~g/L Carbopol microgel is $\tau_y=35\pm5$~Pa and the values of yield stress  obtained with the same method for the other Carbopol microgels used in our experiments are listed in table \ref{Table1} in the main text. 

In order to measure the viscosity curve of the Carbopol microgel post-yield, we performed cyclic rheometric measurements by imposing either torque or angular velocity. We converted angular velocity into shear rate by approximating our rheometric set up to a circular Couette flow geometry. We solved the associated inverse problem for the unknown shear rate function $\dot{\epsilon}(\tau)$ numerically, where the yield stress $\tau_{y}$ was unknown and had to be determined as part of the solution \cite{Yeow:2000,bergemann:2018b}. The resulting values of $\tau_y$ were within $5\%$ of the values measured directly from figure \ref{fig:dynamics_chocolate_carbopol_3}. The benefit of this method is that it does not require any prior assumption about a specific rheological model.

\begin{figure}[!hbt]
\includegraphics[]{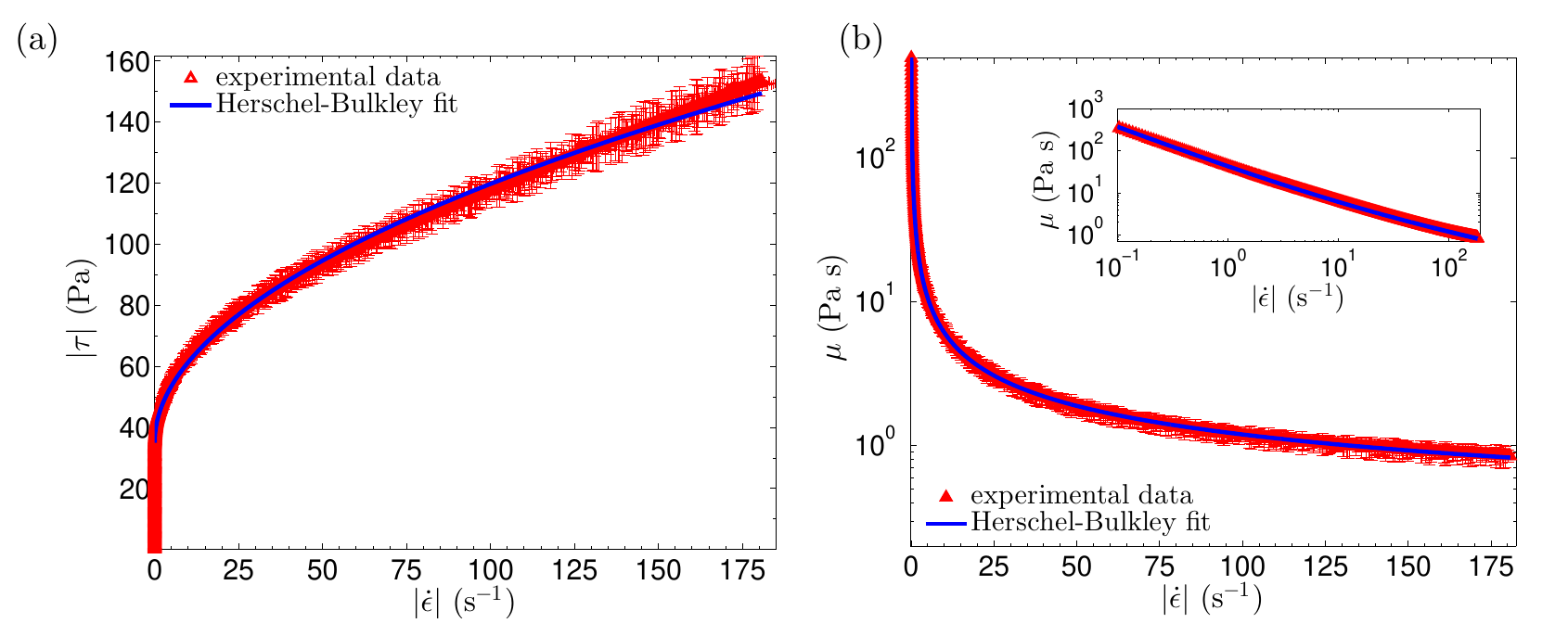}	
	\caption{\small (a) Constitutive behaviour of Carbopol microgel (2.2~g/L) during constant rate ramp-down of either torque or angular velocity. The data is averaged over 4 experiments of 5 cycles and the error bars indicate the standard deviation from the mean. Ramping rates were applied in the range $0.83 \le \alpha \le 2.67$~Pa/s. Variation with shear rate of (a) the shear stress, and (b) the viscosity. The blue lines indicate a two-parameter fit to the data of the Herschel-Bulkley model $\tau = \tau_y +K|\dot{\epsilon}|^n$ fit, using the value of $\tau_y$ previously determined. Inset: Log–log plot of the same graph.}  
	\label{fig:viscosity_measurement}
\end{figure} 

Figure \ref{fig:viscosity_measurement}(a) shows that shear stress increases monotonically as a function of shear rate. Each symbol corresponds to the mean of 4 repetitions of ramp-down experiments (including 5 consecutive cycles) and the error bars indicate the standard deviation from the mean. As in figure \ref{fig:dynamics_chocolate_carbopol_3}(b), the data is insensitive to different ramping rates applied in the range $0.83 \le \alpha \le 2.67$~Pa/s.
The Carbopol microgel is fluid during ramp-down and its flow curve of shear stress as a function of shear rate is accurately captured by a Herschel-Bulkley model, consistent with the literature  \citep{putz:2009,varges:2019}. Carbopol is shear-thinning as indicated by the monotonic decrease of its viscosity $\displaystyle \mu = \tau/\dot{\epsilon}$ with increasing shear rate shown in figure \ref{fig:viscosity_measurement}(b). The straight line on the log-log plot in the inset indicates the power-law behaviour of the Herschel-Bulkley model. A two-parameter fit to the Herschel-Bulkley model gave consistency and shear incidices of $K=8.2$~Pa~s$^n$ and $n=0.49 \pm 0.02$ for 2.2~g/L Carbopol microgel. Results for different concentrations of the Carbopol microgels are listed in table \ref{Table1} in the main text.

\bibliographystyle{abbrvnat}
\bibliography{exp_paper}
\end{document}